\documentclass[10pt]{article}
\usepackage[square,authoryear]{natbib}
\usepackage{graphicx}
\usepackage{marsden_article}
\usepackage{stackrel}

\usepackage[all]{xy}

\usepackage{epstopdf,framed}
\usepackage{pdfsync}
\usepackage{amscd}
\usepackage{ascmac}

\usepackage{mathrsfs}
\usepackage{amscd}
\usepackage{ascmac}

\begin{document}
\newtheorem{remark}[theorem]{Remark}

\title{Dirac structures in nonequilibrium thermodynamics}
\vspace{-0.2in}

\newcommand{\todoFGB}[1]{\vspace{5 mm}\par \noindent
\framebox{\begin{minipage}[c]{0.95 \textwidth} \color{red}FGB: \tt #1
\end{minipage}}\vspace{5 mm}\par}


\author{\hspace{-1cm}
\begin{tabular}{cc}
Fran\c{c}ois Gay-Balmaz &
Hiroaki Yoshimura
\\  CNRS, LMD, IPSL & School of Science and Engineering
\\ Ecole Normale Sup\'erieure & Waseda University
\\  24 Rue Lhomond 75005 Paris, France & Okubo, Shinjuku, Tokyo 169-8555, Japan \\ gaybalma@lmd.ens.fr & yoshimura@waseda.jp\\
\end{tabular}\\\\
}

\maketitle
\vspace{-0.3in}

\begin{center}
\abstract{Dirac structures are geometric objects that generalize both Poisson structures and presymplectic structures on manifolds. They naturally appear in the formulation of constrained mechanical systems.
In this paper, we show that the evolution equations for nonequilibrium thermodynamics admit an intrinsic formulation in terms of Dirac structures, both on the Lagrangian and the Hamiltonian settings.
In absence of irreversible processes these Dirac structures reduce to canonical Dirac structures associated to canonical symplectic forms on phase spaces. Our geometric formulation of nonequilibrium thermodynamic thus consistently extends the geometric formulation of mechanics, to which it reduces in absence of irreversible processes.
The Dirac structures are associated to the variational formulation of nonequilibrium thermodynamics developed in \cite{GBYo2016a,GBYo2016b} and are induced from a nonlinear nonholonomic constraint given by the expression of the entropy production of the system.}
\vspace{2mm}
\color{black}

\end{center}
\tableofcontents

\section{Introduction}\label{Sec_1}

Nonequilibrium thermodynamics is a phenomenological theory which aims to identify and
describe the relations among the observed macroscopic properties of a physical system and to
determine the macroscopic dynamics of this system with the help of fundamental laws
(e.g. \cite{StSc1974}). The field of nonequilibrium thermodynamics naturally
includes macroscopic disciplines such as classical mechanics, fluid dynamics, elasticity, and electromagnetism. The main feature of nonequilibrium thermodynamics is the occurrence of the various irreversible processes, such as friction, mass transfer, and chemical reactions, which are responsible of entropy production.

It is well known that in absence of irreversible processes, the equations of motion of classical mechanics, i.e., the Euler-Lagrange equations, can be derived from Hamilton's variational principle applied to the
action functional associated to the Lagrangian of the mechanical system. This variational formulation is intimately related to the existence of fundamental geometric structures in mechanics such as the canonical symplectic form on phase space, relative to which the equations of motion can be written in Hamiltonian form. It has been a challenging question to systematically extend both the \textit{variational} and the \textit{geometric} structures from the setting of classical mechanics to that of nonequilibrium thermodynamics.

In \cite{GBYo2016a,GBYo2016b}, we proposed an answer to the question regarding the variational structures by presenting a variational formulation for nonequilibrium thermodynamics which extends Hamilton's principle of classical mechanics to include irreversible processes. This approach was developed for discrete and continuum systems with various irreversible processes such as heat transfer, matter transfer, viscosity, and chemical reactions. The variational formulation involves two types of constraints: a \textit{phenomenological constraint} that needs to be satisfied by the critical curve of the action functional; and a \textit{variational constraint} that imposes conditions on the variations of the curve to be considered.

\medskip

\textit{In this paper, we shall show that Dirac structures provide appropriate geometric objects underlying nonequilibrium thermodynamics. These structures are associated to the variational formulation developed in \cite{GBYo2016a,GBYo2016b}.}

\medskip

Dirac structures are geometric objects that generalize both (almost) Poisson structures and (pre)sym\-plectic structures on manifolds. They were originally developed by \cite{CoWe1988}, \cite{Cour1990}, and \cite{Dorfman1993}, who also considered the associated constrained dynamical systems.
Dirac structures were named after Dirac's theory of constraints, see, \cite{Dirac1950}. 
It was shown that Dirac structures are appropriate geometric objects for the formulation of electric circuits, \cite{VaMa1995a,BC1997}, and nonholonomic mechanical systems, \cite{VaMa1995b}, in the context of {\it implicit Hamiltonian systems}, 
which yield implicit differential-algebraic equations as the evolution equations. It was further shown in \cite{YoMa2006a,YoMa2006b} that, in the Lagrangian setting, the Lagrange-d'Alembert equations of nonholonomic mechanics and its associated variational structure, can be formulated in terms of Dirac structures, as implicit Lagrangian systems.

\medskip

The geometry of \textit{equilibrium} thermodynamics has been mainly studied via contact geometry, following the initial works of \cite{Gibbs1873a,Gibbs1873b} and \cite{Ca1909}, by \cite{He1973} and further developments by \cite{Mr1978,Mr1980}, \cite{MrNuScSa1991}.
The contact manifold in this setting is called the \textit{thermodynamic phase space} with contact form given by the Gibbs form, typically, $ \theta = dx ^0- p _i dx^i $ where $x^0$ denotes the energy and  $(x ^i , p _i )$ are pairs of conjugated extensive and intensive variables. In this geometric setting, thermodynamic properties are encoded by Legendre submanifolds of the thermodynamic phase space. A step towards a geometric formulation of irreversible processes was made in \cite{EbMaVa2007} by lifting port Hamiltonian systems to the thermodynamic phase space. The underlying geometric structure in this construction is again a contact form.

\medskip

As mentioned above, in this paper we shall show that the equations of motion for nonequilibrium thermodynamics can be formulated in the context of Dirac structures that are associated with the Lagrangian variational setting developed in \cite{GBYo2016a,GBYo2016b}. More precisely, we shall prove that the equations of motion can be naturally formulated as \textit{Dirac dynamical systems} based either on the generalized energy, the Lagrangian, or the Hamiltonian. To do this, we will extend the use of Dirac structures from the case of linear nonholonomic constraints to a class of nonlinear nonholonomic constraints which we will call \textit{of the thermodynamic type}. In the case of nonequilibrium thermodynamics, the nonlinear constraint is given by the expression of entropy production associated to the irreversible processes involved in the system.

\medskip

\paragraph{Variational formulation for nonequilibrium thermodynamics of simple systems.}
A \textit{discrete thermodynamic system} $ \boldsymbol{\Sigma} $ is a collection $ \boldsymbol{\Sigma} = \cup_{A=1}^N$ of a finite number of interacting simple systems $ \boldsymbol{\Sigma} _A $. By definition, a \textit{simple thermodynamic system}\footnote{In \cite{StSc1974} they are called \'el\'ement de syst\`eme (French). We choose to use the English terminology simple system instead of system element.} is a macroscopic system for which one (scalar) thermal variable and a finite set of mechanical variables are sufficient to describe entirely the state of the system. From the second law of thermodynamics, we can always choose the thermal variable as the entropy $S$, see \cite{StSc1974}. A system is said to be \textit{adiabatically closed} if there is no exchange of matter and heat with the exterior of the system. It is said to be \textit{isolated} if, in addition, there is no exchange of mechanical work with the exterior of the system. 

We now quickly review from \cite{GBYo2016a} the variational formulation of nonequilibrium thermodynamics in the particular case of simple adiabatically closed systems, which are the focus of the present paper. Let $Q$ be the configuration manifold associated to the mechanical variables of the system, assumed to be of finite dimensions. The Lagrangian is a function $L=L(q, \dot q, S): TQ \times \mathbb{R}  \rightarrow \mathbb{R}  $, where $TQ$ denotes the tangent bundle of $Q$. We assume that the system involves exterior and friction forces which are fiber preserving maps $F^{\rm ext}, F^{\rm fr}:TQ\times \mathbb{R}  \rightarrow T^* Q$.

A curve $(q(t),S(t)) \in Q \times \mathbb{R}$, $t \in [t _1 , t _2 ] \subset \mathbb{R}$ is a {\it solution of the variational formulation of nonequilibrium thermodynamics} if it satisfies the variational condition 
\begin{equation}\label{LdA_thermo} 
\delta \int_{t _1 }^{ t _2}L(q , \dot q , S)dt +\int_{t_1}^{t_2}\left\langle F^{\rm ext}(q, \dot q, S), \delta q\right\rangle dt =0, \quad\;\;\; \textsc{Variational Condition}
\end{equation}
for all variations $ \delta q(t) $ and $\delta S(t)$ subject to the constraint
\begin{equation}\label{CV} 
\frac{\partial L}{\partial S}(q, \dot q, S)\delta S= \left\langle F^{\rm fr}(q , \dot q , S),\delta q \right\rangle,\qquad\qquad\, \textsc{Variational Constraint}
\end{equation}
with $ \delta q(t_1)=\delta q(t_2)=0$, and if it satisfies the nonlinear nonholonomic constraint 
\begin{equation}\label{CK} 
\frac{\partial L}{\partial S}(q, \dot q, S)\dot S  = \left\langle F^{\rm fr}(q, \dot q, S) , \dot q \right\rangle. \quad \textsc{Phenomenological Constraint}
\end{equation}
From this variational formulation, we deduce the equations of motion for the adiabatically closed simple system as
\begin{equation}\label{simple_systems} 
\left\{
\begin{array}{l}
\displaystyle\vspace{0.2cm}\frac{d}{dt}\frac{\partial L}{\partial \dot q}- \frac{\partial L}{\partial q}= F^{\rm ext}(q, \dot q, S)+ F^{\rm fr}(q, \dot q, S)\\
\displaystyle\frac{\partial L}{\partial S}\dot S= \left\langle F^{\rm fr}(q, \dot q, S), \dot q \right\rangle .
\end{array} \right.
\end{equation} 

The variational formulation \eqref{LdA_thermo}--\eqref{CK} is an extension of the Hamilton principle of mechanics to the case of nonequilibrium thermodynamics.
In absence of the entropy variable $S$, this variational formulation recovers the Hamilton principle and \eqref{simple_systems} reduces to the Euler-Lagrange equations with external force.

Note that the explicit expression of the constraint \eqref{CK} involves phenomenological laws for the friction
force $F^{\rm fr}$, this is why we refer to it as a \textit{phenomenological constraint}. The associated constraint \eqref{CV}  is called a \textit{variational
constraint} since it is a condition on the variations to be used in \eqref{LdA_thermo}. The constraint \eqref{CK} is nonlinear and one passes from the variational constraint to the phenomenological constraint by formally replacing the variations $ \delta q$, $\delta S$, by the time derivatives $ \dot q$, $\dot S$. Such a simple correspondence between the phenomenological and variational constraints still holds for the more general thermodynamic systems considered in \cite{GBYo2016a,GBYo2016b}.

If the simple system is not adiabatically closed, then the phenomenological constraint \eqref{CK} becomes\begin{equation}\label{Pext} 
\displaystyle\frac{\partial L}{\partial S}\dot S= \left\langle F^{\rm fr}(q, \dot q, S), \dot q \right\rangle -P^{\rm ext} _H
\end{equation} 
where $P^{\rm ext}_H$ denotes the external heat power supply, see \cite{GBYo2016a}.

In the present paper we shall focus on the case of simple isolated systems, i.e., adiabatically closed simple systems in which $F^{\rm ext}=0$. The case $F^{\rm ext}\neq 0$ can be easily included in the Dirac formulation, see Remark \ref{Rmk_Fext}. From now on, we shall refer to these systems as \textit{simple systems}.

Since the partial derivative of the Lagrangian is interpreted as minus the temperature of the system, we shall always assume
\begin{equation}\label{temperature_assumption} 
\frac{\partial L}{\partial S}(q, v, S) < 0.
\end{equation} 

\medskip

It should be noted that the system \eqref{simple_systems} (with the external power included as in \eqref{Pext}) is the general form of the evolution equation for all simple systems, i.e., systems in which one scalar thermal variable is sufficient, in addition to configurational variables, to describe entirely the system. 
The system \eqref{simple_systems} also covers the case of simple systems arising in chemical reactions and electric circuits, see \cite{GBYo2016a} and \S\ref{Examples}. 
The variational formulation \eqref{LdA_thermo}--\eqref{CK}  has been extended to general discrete systems in \cite{GBYo2016a} which involve internal mass and heat transfer, as well as to continuum systems such as multicomponent viscous heat conducting fluid with chemical reactions in \cite{GBYo2016b}, and atmospheric thermodynamics in \cite{GB2017}. We refer to \cite{GBYo2017b} for a variational formulation based on the free energy. For non simple systems, the variational approach is based on the concept of \textit{thermodynamic displacement}.

\begin{remark}[Inclusion of linear nonholonomic mechanical constraints]{\rm Linear nonholonomic \textit{mechanical constraints} can be naturally included in this variational formulation. Suppose that the mechanical motion is constrained by a regular distribution (i.e., a smooth vector subbundle) $ \Delta _Q \subset TQ$. The variational formalism \eqref{LdA_thermo}--\eqref{CK} is modified by considering, in addition to the variational constraint \eqref{CV}, the variational constraint $ \delta q\in \Delta _Q$ and, in addition to the phenomenological constraint \eqref{CK}, the nonholonomic constraint  $\dot  q\in \Delta _Q$. This is a thermodynamic extension of the Lagrange-d'Alembert principle used in nonholonomic mechanics, see, e.g., \cite{Bl2003}.
}
\end{remark}

\begin{figure}[h]
\begin{center}
\includegraphics[scale=0.77]{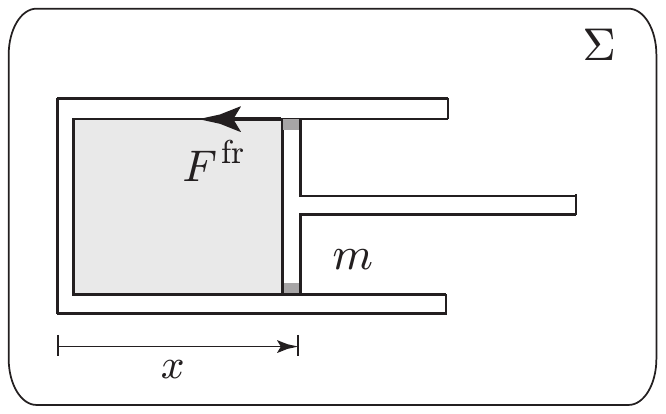}
\caption{Gas confined by a piston}
\label{one_cylinder}
\end{center}
\end{figure}

\paragraph{Example.}
A typical example of a simple thermodynamic system is the case of a perfect gas confined by a piston in a cylinder as illustrated in Fig. \ref{one_cylinder}.
In this case the configuration space of the mechanical variable is $Q= \mathbb{R}  \ni x$ and the Lagrangian $L: T \mathbb{R}  \times \mathbb{R} \rightarrow \mathbb{R}    $ is given by $L(x, \dot x, S) = \frac{1}{2} m\dot x ^2  -U(x, S)$, where $m$ is the mass of the piston, $U(x, S) :=
U(S, V = Ax, N_0)$\footnote{The state functions for a perfect gas are $U=cNRT$ and $pV=NRT$,
where $c$ is a constant depending exclusively on the gas (e.g. $c=\frac{3}{2}$ for monoatomic gas, $c=\frac{5}{2}$ for diatomic gas) and $R$ is the universal gas constant. From this, it is deduced that the internal energy reads
\[
U(S,N,V)= U _0 e^ {\frac{1}{cR}\left(\frac{S}{N}- \frac{S _0 }{N _0 }   \right)  }\left( \frac{N}{N_0}\right) ^{ \frac{1}{c}+1} \left( \frac{V _0 }{V}\right) ^{\frac{1}{c}}.
\]} is the internal energy of the gas, $N_0$ is the number of moles, $V = Ax$ is the volume, and $A$ is the area of the cylinder with constant value.
The friction force reads $F^{\rm fr}(x, \dot x, S)=- \lambda (x, S) \dot x$, where $ \lambda (x, S)\geq 0$ is the phenomenological coefficient,  determined experimentally.
From \eqref{simple_systems}, we immediately get the coupled mechanical and thermal evolution equations for this system as
\begin{equation}\label{equation_piston} 
m\ddot x =p(x,S)A- \lambda (x, S)\dot x, \qquad T(x,S)\dot S= \lambda (x,  S)\dot x ^2,
\end{equation} 
where $p(x,S)=- \frac{\partial U}{\partial V}(x,S)$ is the pressure and $T(x,S)= \frac{\partial U}{\partial S}(x,S)$ is the temperature. We refer to \cite{Gr1999} for a derivation of the equations of motion for this system by following the systematic approach of \cite{StSc1974}.

\paragraph{Plan of the paper.} In \S\ref{section_2}, we consider a 
special class of implicit second order differential-algebraic equations with nonlinear constraints arising from the Lagrange-d'Alembert principle. We refer to the constraints in this class as "\textit{constraints of the thermodynamic type}". We show that these equations admit two types of Dirac formulations, one based on the generalized energy, the other one on the Lagrangian. The Dirac structure for the first case is defined on the Pontryagin bundle (the direct sum of the velocity and phase spaces) while the Dirac structure for the second case is defined on the phase space (cotangent bundle).
In \S\ref{LD_formulation}, we make use of these results to develop two Dirac formulations for the thermodynamics of simple systems. Associated to the first Dirac formulation there exists a variational formulation, the Lagrange-d'Alembert-Pontryagin principle, compatible with the variational formulation for simple systems given in \cite{GBYo2016a}. The Dirac structures constructed in this case are induced by the canonical symplectic form on the \textit{thermodynamic} configuration space.
In \S\ref{section_4}, we develop Dirac formulations, with Dirac structures induced from the canonical symplectic form on the \textit{mechanical} configuration space. These Dirac formulations can be written in terms of  the generalized energy, the Lagrangian, or the Hamiltonian. We also explain the relation between the different Dirac formulations obtained. Finally, in \S\ref{Examples} we illustrate the Dirac formulations with examples of simple thermodynamic systems involving irreversible processes associated to friction, mass transfer, and chemical reactions.

\medskip

We finish this introduction by recalling below the definition of Dirac structures on manifolds and the associated Dirac dynamical systems. We also comment on the role of two canonical symplectic forms: one is associated to the mechanical configuration space and the other to the thermodynamic configuration space.

\paragraph{Dirac structures, Dirac dynamical systems, and nonholonomic mechanics.}
Let $M$ be a manifold and consider the Pontryagin vector bundle $TM \oplus T^*M$ over $M$ endowed with the symmetric fiberwise bilinear form
\[
\left\langle \!\left\langle ( u_m, \alpha _m), (v_m, \beta _m) \right\rangle \! \right\rangle = \left\langle\beta _m, u_m\right\rangle + \left\langle \alpha _m, v_m \right\rangle ,
\]
for $( u_m, \alpha _m), (v_m, \beta _m) \in T_mM\oplus T^*_mM$.
A \textit{Dirac structure on  $M$} (also called an almost Dirac structure) is by definition a vector subbundle $D \subset TM \oplus T^*M$, such that $D^\perp=D$ relative to $\left\langle \!\left\langle \cdot ,\cdot \right\rangle \! \right\rangle $, see \cite{Cour1990}.

For example, given a two-form $ \omega \in \Omega ^2 (M)$ and a distribution $ \Delta_M $ on $M$ (i.e., a vector subbundle of $TM$), the subbundle $D_{ \Delta _M} \subset TM \oplus T^*M$ defined by, for each $m \in M$,
\begin{equation}\label{def_Dirac}
\begin{aligned}
D_{ \Delta _M}(m) :
&=\big\{ (v_{m}, \alpha_{m}) \in T_{m}M  \times T^{\ast}_{m}M  \; \mid \;  v_{m} \in \Delta_M(m) \; \text{and} \\ 
& \qquad \qquad \qquad \left\langle \alpha_{m},w_{m} \right\rangle =\omega(m)(v_{m},w_{m}) \; \;
\mbox{for all} \; \; w_{m} \in  \Delta _M(m) \big\}
\end{aligned}
\end{equation}
is a Dirac structure on $M$. In this paper, we shall extensively use this construction of Dirac structure. 

\medskip

Given a Dirac structure $D$ on $M$ and a function $ F: M \rightarrow \mathbb{R}$, the associated \textit{Dirac dynamical system} for a curve $m(t) \in M$ is
\begin{equation}\label{DDS}
\big(\dot m(t), \mathbf{d} F(m(t)) \big) \in D(m(t)).
\end{equation} 
Note that in general \eqref{DDS} is a system of implicit differential-algebraic equations and the questions of existence, uniqueness or extension of solutions for a given initial condition can present several difficulties.

The formulation \eqref{DDS} offers a unified treatment for the equations of nonholonomic mechanics with (possibly degenerate) Lagrangian. Let us quickly recall how this proceeds. Consider a mechanical system with configuration manifold $Q$, a Lagrangian $L:TQ \rightarrow \mathbb{R}  $ and a constraint distribution $ \Delta_Q \subset TQ$. Let $P:= TQ \oplus T^*Q$ be the Pontryagin bundle of $Q$ and define the induced distribution $ \Delta _P(q,v,p):= \left( T_{(q,v,p)} \pi _{(P, Q)} \right) ^{-1} ( \Delta _Q(q))$, where $ \pi _{(P,Q)}: P \rightarrow Q$ is the projection defined by $ \pi _{(P,Q)}(q,v,p)=q$. We consider the Dirac structure $D_{ \Delta _P}$ induced, via \eqref{def_Dirac},  by the distribution $ \Delta _P$ and the presymplectic form $ \omega _P:= \pi _{(P, T^*Q)} ^\ast\Omega _{T^*Q}$ on $P$, where $\Omega _{T^*Q}$ is the canonical symplectic form on $ T^*Q$ and $ \pi _{(P, T^*Q)}: P \rightarrow T^*Q$ is the projection defined by $\pi _{(P, T^*Q)}(q,v,p)= (q,p)$. Then, the equations of motion for the nonholonomic system can be written in the form of the {\bfi Dirac dynamical system on the Pontryagin bundle} as
\begin{equation}\label{DDS_NH} 
\big((\dot q(t), \dot v(t), \dot p(t)) ,\mathbf{d} E(q(t), v(t), p(t)) \big) \in D_{\Delta _P}(q(t), v(t), p(t)),
\end{equation} 
where $E: P \rightarrow \mathbb{R}  $, defined by $E(q,v,p)= \left\langle p, v\right\rangle - L(q,v)$ is the \textit{generalized energy} associated to $L$. System \eqref{DDS_NH} is equivalent to the implicit differential-algebraic equation
\begin{equation}\label{implicit_NH} 
\left\{
\begin{array}{l}
\displaystyle \vspace{0.2cm} \dot q(t) \in \Delta _Q(q(t)), \qquad v(t)=\dot q(t),\\
\displaystyle \vspace{0.2cm}p(t) = \frac{\partial L}{\partial v} (q(t),v(t)), \qquad  \dot p(t)-  \frac{\partial L}{\partial q}(q(t),v(t)) \in  \Delta _Q(q(t))^\circ,
\end{array} \right.
\end{equation} 
where $ \Delta _Q(q) ^ \circ$ denotes the annihilator of $\Delta _Q(q)$ in $T^*_qQ$.
This system implies the Lagrange-d'Alembert equations
\[
\frac{d}{dt} \frac{\partial L}{\partial v} (q(t),v(t))- \frac{\partial L}{\partial q}(q(t),v(t)) \in  \Delta _Q(q(t))^\circ, \qquad \dot q(t) \in \Delta _Q(q(t))
\]
for the nonholonomic system, see \cite{YoMa2006b}.

There is an alternative Dirac formulation of the implicit system \eqref{implicit_NH} that makes use of the Dirac structure $D_{ \Delta _{T^*Q}}$ on $T^* Q $ (instead of $P$) associated to the canonical symplectic form $ \Omega _{T^*Q}$ on $T^*Q$ and to the distribution $ \Delta _{T^*Q}$ induced on $T^* Q$ by $ \Delta _Q$. This formulation is directly based on the Lagrangian function $L:TQ \rightarrow \mathbb{R}  $ and reads
\begin{equation}\label{LD_NH} 
\big((\dot q(t), \dot p(t)) ,\mathbf{d} _D L(q(t), v(t)) \big) \in D_{\Delta _{T^*Q}}(q(t), p(t)),
\end{equation} 
where $ \mathbf{d} _DL$ is the Dirac differential whose definition will be reviewed later. The system \eqref{LD_NH} is called the {\bfi Lagrange-Dirac system}.

In the hyperregular case, the equations can be also written in terms of the Hamiltonian $H: T^*Q \rightarrow \mathbb{R}  $ in the {\bfi Hamilton-Dirac} form
\begin{equation}\label{HD_NH} 
\big((\dot q(t), \dot p(t)) ,\mathbf{d} H(q(t), p(t)) \big) \in D_{\Delta _{T^*Q}}(q(t), p(t)).
\end{equation} 

In the present paper we shall show that the equations of evolution for the thermodynamics of simple systems can be written in the Dirac dynamical system form \eqref{DDS_NH} and in the Lagrange-Dirac form \eqref{LD_NH} in two ways, as well as in the Hamilton-Dirac form \eqref{HD_NH}, see Fig. \ref{big_diagram}. In the case of thermodynamics, however, one starts with a nonlinear constraint.
We shall restrict our discussion to the case of simple thermodynamic systems and leave the development of the corresponding Dirac approach to general thermodynamic systems for a future work.

\paragraph{Mechanical and thermodynamic configuration spaces.} As mentioned earlier, in absence of the entropy variable, both constraints disappear and the variational formulation \eqref{LdA_thermo}--\eqref{CK} recovers Hamilton's principle of classical mechanics. In this case the intrinsic geometric structure 
is usually given by the canonical symplectic form $ \Omega _{T^*Q}$ on the cotangent bundle of the configuration manifold $Q$ of the mechanical system, locally given as
\[
\Omega _{T^*Q}= dq ^i \wedge dp _i.
\]
It is therefore expected that when thermodynamics is included, the relevant geometric structure should be naturally induced from this canonical structure. We shall show in this paper that this is indeed the case, namely, the appropriate Dirac structures are induced by the canonical symplectic form $ \Omega _{T^*Q}$ and by a distribution built from the phenomenological constraint. These Dirac structures are developed on $M=(TQ \oplus _{Q \times \mathbb{R}  }T^* Q)$  and $ N=T^* Q \times \mathbb{R} $ for the generalized energy and the Lagrangian formulations, respectively.

In addition, we will see that the canonical symplectic structure on the \textit{thermodynamic configuration space}
\[
\mathcal{Q} := Q \times \mathbb{R}  \ni (q,S),
\]
locally given by
\[
\Omega _{T^* \mathcal{Q} }= dq ^i \wedge dp _i+ dS \wedge  d\Lambda ,
\]
can be also used as the fundamental intrinsic form from which Dirac structures are deduced.

\section{A class of constraints of the thermodynamic type and its Dirac formulations}\label{section_2}

In this section, we first recall a generalized Lagrange-d'Alembert principle for nonlinear nonholonomic mechanics, which involves two types of constraints, the \textit{kinematic} and \textit{variational constraints}. Then we focus on a specific situation, relevant with thermodynamics, for which \textit{the kinematic constraint is deduced from the variational constraint}. We show that in this specific case, which we call ``\textit{constraints of the thermodynamic type}", the equations of motion derived from the Lagrange-d'Alembert principle admit two types of Dirac formulations; one associated to a Dirac structure on the cotangent bundle and the other one associated to a Dirac structure on the Pontryagin bundle.

\subsection{Variational formulation for nonlinear nonholonomic systems}\label{VF_NonlinearNH} 

\paragraph{The Lagrange-d'Alembert principle.} Consider a configuration manifold $Q$ and a Lagrangian function $L:TQ \rightarrow \mathbb{R}$ defined on the tangent bundle $TQ$ of the configuration manifold. The general formulation of the Lagrange-d'Alembert principle, considered in \cite{CeIbdLdD2004}, involves the choice of two distinct constraints, the kinematical and variational constraints, which, following \cite{Ma1998} must in general be considered as independent notions.
The kinematical constraint is given by a submanifold $C_K \subset TQ$ and describes a restriction on the motion of the mechanical system. The variational constraint is a subset $C_V \subset TQ \times _Q TQ$ such that for all $(q,v)\in T_qQ$ the sets
\[
C_V(q,v):=C_V\cap \left(  \{(q,v)\} \times T_qQ \right)
\]
are vector spaces, the spaces of virtual displacements.

For simplicity, we assume that $\pi_{(TQ,Q)}(C_K)=Q$, where $\pi_{(TQ,Q)}:TQ \rightarrow Q$ is the tangent bundle projection, we assume that $C_V$ is a submanifold and also that all the vector spaces $C_V(q,v)$ have the same (strictly positive) dimension. These hypotheses are not needed to apply the Lagrange-d'Alembert principle, however, since they are verified in the case of interest for this paper, we shall assume them. We also refer to \cite{CeGr2007} for a general treatment of the constraint sets $C_V$ and $C_K$.

By definition, a curve $q:[t _1 , t _2 ] \rightarrow Q$ is a solution of the Lagrange-d'Alembert principle for $(L, C_K, C_V)$ if $(q(t),\dot q(t)) \in C_K$ and
\[
\left.\frac{d}{d\varepsilon}\right|_{\varepsilon=0}  \int_{t_1 }^{t_2 }L(q_ \varepsilon (t), \dot q_ \varepsilon (t)) dt=0, 
\]
for all variations $ q_ \varepsilon (t)$ of the curve $q(t)$ with fixed endpoints, such that
\[
\delta q(t) :=\left.\frac{d}{d\varepsilon}\right|_{\varepsilon=0} q_ \varepsilon (t) \in C_V(q(t), \dot q(t)).
\]
By applying this principle, one obtains that a curve $q:[t _1 , t _2 ] \rightarrow Q$ is a solution of the Lagrange-d'Alembert principle for $(L, C_K, C_V)$ if and only if it  satisfies the following implicit second order differential-algebraic equations
\begin{equation}\label{2nd_order} 
(q(t),  \dot q(t)) \in C_K\quad\text{and}\quad \frac{d}{dt} \frac{\partial L}{\partial \dot q}- \frac{\partial L}{\partial q} \in C_V(q(t), \dot q(t))^\circ.
\end{equation} 
We shall not discuss here the question of the existence and uniqueness of solutions of this implicit differential equation since in our case of interest for thermodynamics, see \S\ref{LD_formulation}, it reduces to an explicit differential equation.

\paragraph{Nonlinear constraints of the thermodynamic type.} We now present a setting that is well adapted for the description of the nonequilibrium thermodynamics of simple systems. From now on, we shall denote the configuration manifold by $ \mathcal{Q} $, while we use the notation $Q$ exclusively for the configuration manifold associated to the mechanical variables. As we will see in \S\ref{LD_formulation} when applying this setting to thermodynamics, the configuration manifold $ \mathcal{Q} $ contains mechanical variables as well as thermodynamic variables, so it is referred to as the \textit{thermodynamic configuration space}. 

In the setting that we now develop, the variational constraint $C_{V} \subset T \mathcal{Q} \times _ \mathcal{Q} T \mathcal{Q} $ is chosen first and the kinematic constraint is defined from it as follows:
\begin{equation}\label{def_CK}
C_{K}:=\{ (q, v) \in  T \mathcal{Q}  \mid (q, v)  \in C_{V}(q, v)\}.
\end{equation} 
Thus, if $C_{V}$ is locally given by equations $ \Phi (q,v)_i ^a    \delta q ^i =0$, then $C_{K}$ is locally given by $ \Phi (q, v)_i ^a   v ^i=0 $.

Note that in this setting, it is the variational constraint which determines the kinematic constraint, and not the converse as in the Chetaev case. Note also that in general $C_{K} \subset T \mathcal{Q} $ is a {\it nonlinear nonholonomic constraint}.

In the case of a kinematic constraint given by \eqref{def_CK}, the implicit second order differential-algebraic equation \eqref{2nd_order} becomes
\begin{equation}\label{2nd_order_therm} 
(q(t),  \dot q(t)) \in C_{V}(q(t),  \dot q(t)) \quad\text{and}\quad \frac{d}{dt} \frac{\partial L}{\partial \dot q}- \frac{\partial L}{\partial q} \in C_V(q(t), \dot q(t))^\circ,
\end{equation} 
where we note that $(q(t),  \dot q(t)) \in C_{V}(q(t),  \dot q(t))$ is equivalent with $(q(t),  \dot q(t)) \in C_{K}$.
One directly computes that the energy $E:T \mathcal{Q}  \rightarrow \mathbb{R}  $, defined by
\[
E(q,v):= \left\langle \frac{\partial L}{\partial v}(q,v), v \right\rangle - L(q,v)
\]
is constant along the solutions of the system \eqref{2nd_order_therm}, for any choice of variational constraint $C_{V}$.

We shall present two Dirac formulations for the implicit second order differential equations \eqref{2nd_order}. These two formulations, on the Pontryagin bundle $T \mathcal{Q}  \oplus T^* \mathcal{Q} $ and on the cotangent bundle $T^* \mathcal{Q} $, extend the corresponding Dirac formulations \eqref{DDS_NH} and \eqref{LD_NH} of nonholonomic mechanics, respectively. While the first formulation uses the generalized energy and its ordinary differential, the second formulation uses the Lagrangian and its Dirac differential.

\subsection{Dirac formulation on the Pontryagin bundle}\label{LD_Pontryagin}

In this section we develop the Dirac dynamical system formulation on the Pontryagin bundle $ \mathcal{P} =T \mathcal{Q}  \oplus T^{\ast} \mathcal{Q} $ for systems with nonlinear nonholonomic constraints of the thermodynamic type. The Dirac structure is defined from the presymplectic form on $ \mathcal{P} $ induced by the canonical symplectic form on $T^* \mathcal{Q} $ and from a distribution on $ \mathcal{P} $ induced by the given variational constraint $C_V$. This results in a Dirac dynamical system extending \eqref{DDS_NH} to the case of nonlinear constraints of the thermodynamic type.

\paragraph{Induced Dirac structure.} From a given variational constraint $C_V \subset T \mathcal{Q}  \times _ \mathcal{Q}  T \mathcal{Q} $, we construct the \textit{induced distribution} $\Delta _ \mathcal{P}$ on the Pontryagin bundle $\pi_{( \mathcal{P} , \mathcal{Q} )}: \mathcal{P} =T \mathcal{Q}  \oplus T^* \mathcal{Q} \rightarrow \mathcal{Q} $, defined by:
\begin{equation}\label{def_tilde_CV} 
\Delta _ \mathcal{P} (q,v, p):= \left( T_{(q, v, p)} \pi_{(\mathcal{P} ,\mathcal{Q})} \right) ^{-1} (C_V(q,v))\subset T_{(q, v, p)}\mathcal{P} , \quad \text{for all $(q,v,p) \in \mathcal{P} $}.
\end{equation} 
Locally, the distribution reads
\[
\Delta_ \mathcal{P} (q, v, p)=\big\{(q, v, p, \delta q, \delta v, \delta p) \in T_{(q,v,p)}\mathcal{P}  \mid (q, \delta q) \in C_V(q,v)\big\}.
\]
Consider the presymplectic form $  \omega _\mathcal{P} := \pi_{(\mathcal{P} ,T^{\ast}\mathcal{Q})} ^\ast \Omega_{T^*\mathcal{Q}} $ on $\mathcal{P} $ induced from the canonical symplectic form $ \Omega _{T^*\mathcal{Q}}$ on $T^*\mathcal{Q}$ by the projection $\pi_{(\mathcal{P} ,T^{\ast}\mathcal{Q})}:\mathcal{P} \rightarrow T^*\mathcal{Q}$. To the distribution $\Delta_ \mathcal{P} $ and the presymplectic form $\omega _\mathcal{P} $ is naturally associated 
the \textit{induced Dirac structure} $D_{\Delta_ \mathcal{P} }$ on $\mathcal{P} $ defined as
\begin{equation}\label{D_thermo}
\begin{aligned}
D_{\Delta_ \mathcal{P} }(x) :
&=\big\{ (v_{x}, \alpha_{x}) \in T_{x}\mathcal{P}  \times T^{\ast}_{x}\mathcal{P}  \; \mid \;  v_{x} \in \Delta_ \mathcal{P} (x) \; \text{and} \\ 
& \qquad \qquad \qquad \left\langle \alpha_{x},w_{x} \right\rangle =\omega_{\mathcal{P} }(x)(v_{x},w_{x}) \; \;
\mbox{for all} \; \; w_{x} \in \Delta_ \mathcal{P} (x) \big\},
\end{aligned}
\end{equation}
where we used the notation $x=(q,v,p) \in \mathcal{P} $. Note that this definition follows the construction of a Dirac structure recalled in \eqref{def_Dirac}, with $M= \mathcal{P} $, $ \omega=  \omega _{ \mathcal{P} }$ and $\Delta = \Delta_ \mathcal{P} $. Locally, one directly checks that $\big((q,v,p,\dot q, \dot v, \dot p),(q,v,p, \alpha , \beta , u)\big) \in D_{ \Delta_ \mathcal{P} }(q,v,p)$ is equivalent to
\begin{equation}\label{local}  
(q, \dot q) \in C_V(q,v), \quad \beta =0 , \quad u=\dot q, \quad \dot p + \alpha  \in C_V(q,v)^\circ.
\end{equation}

\paragraph{Dirac dynamical system formulation on $\mathcal{P} = T\mathcal{Q} \oplus T^*\mathcal{Q} $.}
Given a Lagrangian function $L:T\mathcal{Q} \rightarrow \mathbb{R}  $, we define the \textit{generalized energy} on the Pontryagin bundle as
\begin{equation}\label{gen_energy} 
\mathcal{E} :  \mathcal{P}  \rightarrow \mathbb{R}  , \quad \mathcal{E} (q,v,p):=\left\langle p, v \right\rangle -L(q,v).
\end{equation} 
Using the local expression $ \mathbf{d} \mathcal{E} (q,v,p)= \big( q,v,p, -\frac{\partial L}{\partial q}, p - \frac{\partial L}{\partial v},v \big )$ for the differential of $ \mathcal{E} $ together with \eqref{local}, it follows that $\big((q,v,p,\dot q, \dot v, \dot p),\mathbf{d}  \mathcal{E} (q,v,p)\big) \in D_{ \widetilde{C}_V}(q, v,p)$ is equivalent to
\begin{equation}\label{DiracTherm}
(q, \dot q) \in C_V(q,v), \quad p - \frac{\partial L}{\partial v} =0 , \quad v=\dot q, \quad \dot p -\frac{\partial L}{\partial q} \in C_V(q,v)^\circ.
\end{equation}
One observes that from the definition of $C_K$ given in \eqref{def_CK} and from the third equation, the first equation is equivalent to $(q, \dot q) \in C_K$. This is a key step that explicitly involves the fact that the kinematic constraint $C_K$ is determined from the variational constraint $C_V$ via the definition \eqref{def_CK}.
Therefore, we get from \eqref{DiracTherm} the implicit form of the equations \eqref{2nd_order_therm} associated to the constraints of the thermodynamic type. These results are summarized in the following Theorem.

\begin{theorem}\label{theorem}  Consider a variational constraint $C_V \subset T \mathcal{Q} \times _ \mathcal{Q}  T\mathcal{Q}$, the associated kinematic constraint $C_K$ of the thermodynamic type defined in \eqref{def_CK}, and the associated Dirac structure $D_{\Delta_ \mathcal{P}}$ on $\mathcal{P} $ defined in \eqref{D_thermo}. Let $L:T\mathcal{Q} \rightarrow \mathbb{R}  $ be a Lagrangian function and $ \mathcal{E} : \mathcal{P} \rightarrow \mathbb{R}  $ be the associated generalized energy. Then the following statements are equivalent:
\begin{itemize}
\item The curve $(q(t),v(t),p(t)) \in \mathcal{P} $ satisfies the implicit first order differential-algebraic equations
\begin{equation}\label{implicit_thermo} 
\left\{
\begin{array}{l}
\displaystyle \vspace{0.2cm}(q(t), \dot q(t)) \in C_K, \qquad v(t)=\dot q(t),\\
\displaystyle \vspace{0.2cm}p(t) = \frac{\partial L}{\partial v} (q(t),v(t)), \qquad  \dot p(t)-  \frac{\partial L}{\partial q}(q(t),v(t)) \in C_V(q(t),v(t))^\circ.
\end{array} \right.
\end{equation} 
\item The curve $(q(t),v(t),p(t)) \in \mathcal{P} $ satisfies the {\bfi Dirac dynamical system}
\[
\big((q,v,p ,\dot q, \dot v , \dot p  ),\mathbf{d}  \mathcal{E} (q  ,v ,p  )\big) \in D_{ \Delta _ \mathcal{P} }(q,v,p ).
\]
\end{itemize}
Moreover, the system \eqref{implicit_thermo} is equivalent to the implicit second order differential-algebraic equations on $\mathcal{Q} $ given by
\[
(q(t),  \dot q(t)) \in C_K\quad\text{and}\quad \frac{d}{dt} \frac{\partial L}{\partial \dot q}- \frac{\partial L}{\partial q} \in C_V(q(t), \dot q(t))^\circ,
\]
associated to constraints of the thermodynamic type.
\end{theorem}

\paragraph{The Lagrange-d'Alembert-Pontryagin  principle.} We now show that the evolution equations \eqref{implicit_thermo} associated to the Dirac dynamical system on the Pontryagin bundle $\mathcal{P} $ have a natural variational structure, called the Lagrange-d'Alembert-Pontryagin principle. 
This is similar to the situation of Dirac dynamical systems in linear nonholonomic mechanics, see \cite{YoMa2006b}, which also admit such variational structures.

For constraints of the thermodynamic type, the \textit{Lagrange-d'Alembert-Pontryagin principle} for a curve $x(t)=(q(t),v(t), p(t)) \in \mathcal{P} =T\mathcal{Q} \oplus T^{\ast}\mathcal{Q}$ in the Pontryagin bundle reads as follows:
\begin{equation}\label{LD_variational_explicit}
\delta \int_{t_1}^{t_2}\Big[\! \left\langle p, \dot q -v\right\rangle +L (q,v)\Big] dt=0,
\end{equation}
for variations $ \delta q(t), \delta v(t), \delta p(t)$ such that $ \delta q(t) \in C_V(q (t) , v (t) )$ and $ \delta q(t_1)= \delta q(t_2)=0$, and where the curve $(q(t),v(t),p(t))$ satisfies $\dot q(t) \in C_V(q(t), v(t))$. In terms of $x(t)=(q(t),v(t), p(t)) \in \mathcal{P} $, these conditions intrinsically read $ \delta x (t)\in \Delta_ \mathcal{P} (x(t))$, $ T\pi _{(P,Q)}(\delta x(t_1))=T\pi _{(P,Q)}(\delta x(t_2))=0$, and $\dot x(t) \in \Delta_ \mathcal{P} (x(t))$. Notice that the variational formulation \eqref{LD_variational_explicit} may be restated in terms of the generalized energy $ \mathcal{E} :\mathcal{P}  \rightarrow \mathbb{R}  $, see \eqref{gen_energy}, as
\begin{equation}\label{LD_variational}
\delta \int_{t_1}^{t_2}\Big[\! \left\langle p, \dot q \right\rangle -\mathcal{E} (q,v,p)\Big] dt=0,
\end{equation} 
with respect to the same class of variations as before. One checks that a curve $x(t)$ satisfies the principle \eqref{LD_variational_explicit} if and only if it is a solution of the equations \eqref{implicit_thermo}, thus called the \textit{Lagrange-d'Alembert-Pontryagin equations} associated to $C_V$ and $L$.

This variational formulation can be intrinsically given as follows. Consider the one-form
$ \theta _\mathcal{P} :=\pi_{( \mathcal{P} ,T^{\ast}\mathcal{Q} )}^\ast\Theta_{T^{\ast} \mathcal{Q} } \in \Omega ^1 (\mathcal{P}  )$, where $\Theta_{T^{\ast}\mathcal{Q} } $ is the canonical one-form on $ T^* \mathcal{Q} $. In coordinates, we have $ \theta _  \mathcal{P} (q,v,p)=p _i dq^i $. With the help of this one-form, the principle \eqref{LD_variational} can be written as
\[
\delta \int_{ t _1 }^{ t _2 } \Big[ \!\left\langle \theta _ \mathcal{P}  (x) , \dot x \right\rangle - \mathcal{E} (x)\Big] dt=0,
\]
with respect to the same class of variations as before. This yields the \textit{intrinsic Lagrange-d'Alembert-Pontryagin equations}:
\[
\mathbf{i} _ {\dot x} \omega _{P }(x) - \mathbf{d} \mathcal{E}(x) \in \Delta_ \mathcal{P} (x)^\circ, \quad \dot x \in \Delta_ \mathcal{P} (x),
\]
where $ \omega _  \mathcal{P}  = - \mathbf{d} \theta _  \mathcal{P} = \pi_{( \mathcal{P} ,T^{\ast}\mathcal{Q} )} ^\ast \Omega _{ T^* \mathcal{Q} }$ is the presymplectic form on $ \mathcal{P} $ considered earlier.

\subsection{Lagrange-Dirac formulation}\label{LD_TsQ}

In this section we develop the Lagrange-Dirac formulation for systems with nonlinear nonholonomic constraints of the thermodynamic type. We define the Dirac structure on $T^* \mathcal{Q} $ from the canonical symplectic form on $T^* \mathcal{Q} $ and from a distribution on $T^*\mathcal{Q}  $ induced by the given variational constraint $C_V$. This results in a Lagrange-Dirac system extending \eqref{LD_NH} to the case of nonlinear constraints of the thermodynamic type.

\paragraph{Induced Dirac structure.} From a given variational constraint $C_V \subset T\mathcal{Q}  \times _\mathcal{Q}  T\mathcal{Q} $ and a given Lagrangian $L:T\mathcal{Q}  \rightarrow \mathbb{R}$, we assume that it is possible to define the variational constraint $ \mathscr{C} _V \subset T^*\mathcal{Q}  \times _\mathcal{Q}  T\mathcal{Q} $ by
\begin{equation}\label{CCC_V} 
\mathscr{C} _V(q,p):= C_V(q,v),
\end{equation} 
where $v$ is such that $\frac{\partial L}{\partial v}(q,v)=p$. The assumption under this definition is that the right hand side of \eqref{CCC_V} does not depend on the choice of $v$ such that $\frac{\partial L}{\partial v}(q,v)=p$ holds. This assumption holds for instance when $L$ is nondegenerate. However, in \S\ref{LD_formulation}, we will show that this construction can be used even for a special case in which $L$ is degenerate.

Using the cotangent bundle projection $\pi_{(T^{\ast}\mathcal{Q} ,\mathcal{Q} )}:T^{\ast}\mathcal{Q}  \to \mathcal{Q} $ we construct from $ \mathscr{C} _V$ the \textit{induced distribution} $\Delta_ {T^*\mathcal{Q} }$ on $T^*\mathcal{Q} $ as
\begin{equation}\label{def_tilde_CCCV} 
\Delta_ {T^*\mathcal{Q} }(q,p):= \left( T_{(q, p)} \pi_{(T^{\ast}\mathcal{Q} ,\mathcal{Q} )} \right) ^{-1} ( \mathscr{C} _V(q,p))\subset T_{(q, p)}T^*\mathcal{Q} , \quad \text{for all $(q,p) \in T^*\mathcal{Q} $}.
\end{equation} 
Locally, the distribution reads
\[
\Delta_ {T^*\mathcal{Q} }(q, p)=\big\{(q, p, \delta q, \delta p) \in T_{(q,p)}T^*\mathcal{Q}  \mid (q, \delta q) \in \mathscr{C} _V(q,p)\big\}.
\]
To the distribution $\Delta_ {T^*\mathcal{Q} }$ and the canonical symplectic form $\Omega_{T^* \mathcal{Q} }$ on $T^*\mathcal{Q} $ is naturally associated the 
\textit{induced Dirac structure} $D_{ \Delta_ {T^*\mathcal{Q} }}$ on $T^*\mathcal{Q} $  defined, for $z=(q,p) \in T^*\mathcal{Q} $, by
\begin{equation}\label{D_thermo_qp}
\begin{aligned}
D_{\Delta_ {T^*\mathcal{Q} }}(z) :
&=\big\{ (v_{z}, \alpha_{z}) \in T_{z}T^*\mathcal{Q}  \times T^{\ast}_{z}T^*\mathcal{Q}  \; \mid \;  v_{z} \in \Delta_ {T^*\mathcal{Q} }(z) \; \text{and} \\ 
& \qquad \qquad \qquad \left\langle \alpha_{z},w_{z} \right\rangle =\Omega_{T^* \mathcal{Q} }(z)(v_{z},w_{z}) \; \;
\mbox{for all} \; \; w_{z} \in \Delta_ {T^*\mathcal{Q} }(z) \big\}.
\end{aligned}
\end{equation}
Locally, one directly checks that $\big((q,p,\dot q,  \dot p),(q,p, \alpha  , u)\big) \in D_{ \Delta _{T^* \mathcal{Q} }}(q,p)$ is equivalent to
\begin{equation}\label{local_LD}  
(q, \dot q) \in \mathscr{C} _V(q,p), \quad u=\dot q, \quad \dot p + \alpha \in \mathscr{C} _V(q,p)^\circ.
\end{equation}

\paragraph{Lagrange-Dirac formulation on $T^{\ast}\mathcal{Q} $.}
Given a Lagrangian function $L:T\mathcal{Q}  \rightarrow \mathbb{R}$, the Dirac differential of the Lagrangian $L$ is the map $\mathbf{d}_DL:T\mathcal{Q}  \rightarrow T^*T^*\mathcal{Q} $ defined by
\[
\mathbf{d}_DL(q,v):=(\gamma _\mathcal{Q}  \circ \mathbf{d} L)(q,v)=  \left( q, \frac{\partial L}{\partial v}, - \frac{\partial L}{\partial q}, v \right),
\]
where $ \gamma _\mathcal{Q} : T^*T\mathcal{Q}  \rightarrow T^*T^*\mathcal{Q} $ is the canonical diffeomorphism, locally given by $ \gamma _\mathcal{Q} (q, v, \alpha  ,p)=(q,p,- \alpha, v)$, see, \cite{YoMa2006a}. Using the local expression \eqref{local_LD}, one checks that $\big((q, p, \dot q, \dot p),  \mathbf{d}_DL( q, v) \big)   \in D_{\Delta_ {T^*\mathcal{Q} }}(q, p)$  is equivalent to
\[
p =\frac{\partial L}{\partial v} , \quad (q, \dot q) \in  \mathscr{C} _V(q,p), \quad  v=\dot q, \quad  \dot p -\frac{\partial L}{\partial q}\in \mathscr{C} _V(q,p)^\circ.
\]
We thus get again the implicit form of the equations \eqref{2nd_order_therm} associated to the nonholonomic constraints of the thermodynamic type. These results are summarized in the following Theorem.

\begin{theorem}\label{theorem_LD}  Consider a variational constraint $C_V \subset T\mathcal{Q}  \times _\mathcal{Q}  T\mathcal{Q} $ and the associated kinematic constraint $C_K$ of the thermodynamic type defined in \eqref{def_CK}. Let  $L:T\mathcal{Q}  \rightarrow \mathbb{R}  $ be a Lagrangian function, $ \mathscr{C} _V \subset T^*\mathcal{Q}  \times _\mathcal{Q}  T\mathcal{Q} $ be the variational constraint  as in \eqref{CCC_V}, and define the induced Dirac structure $D_{\Delta_ {T^*\mathcal{Q} }}$ as in \eqref{D_thermo_qp}. Then the following statements are equivalent:
\begin{itemize}
\item The curve $(q(t),v(t),p(t)) \in\mathcal{P} $ satisfies the implicit first order differential-algebraic equations
\begin{equation}\label{implicit_thermo_qp} 
\left\{
\begin{array}{l}
\displaystyle \vspace{0.2cm}(q(t), \dot q(t)) \in \mathscr{C} _V(q(t),p(t)), \qquad v(t)=\dot q(t),\\
\displaystyle \vspace{0.2cm}p(t) = \frac{\partial L}{\partial v} (q(t),v(t)), \qquad  \dot p(t)-  \frac{\partial L}{\partial q}(q(t),v(t)) \in \mathscr{C} _V(q(t),p(t))^\circ.
\end{array} \right.
\end{equation} 
\item The curve $(q(t),v(t),p(t)) \in \mathcal{P} $ satisfies the {\bfi Lagrange-Dirac system}
\[
\big((q,p ,\dot q ,  \dot p ),\mathbf{d}_DL  (q ,v  )\big) \in D_{ \Delta_ {T^*\mathcal{Q} }}(q,p ).
\]
\end{itemize}
Moreover, the system \eqref{implicit_thermo_qp} is equivalent to the implicit second order differential-algebraic equations on $\mathcal{Q} $ given by
\begin{equation}\label{LdA_NL} 
(q(t),  \dot q(t)) \in C_K\quad\text{and}\quad \frac{d}{dt} \frac{\partial L}{\partial \dot q}- \frac{\partial L}{\partial q} \in C_V(q(t), \dot q(t))^\circ
\end{equation} 
associated to nonlinear nonholonomic constraints of the thermodynamic type.
\end{theorem}

\begin{remark}[The special case of mechanical systems with linear nonholonomic constraints]{\rm In the particular case of linear nonholonomic constraints in mechanics, the previous two Dirac formulations recover those of nonholonomic mechanical systems developed in \cite{YoMa2006a}. In this case $ \mathcal{Q}  =Q$ and we assume that a Lagrangian $L:TQ \rightarrow \mathbb{R}  $ and a distribution $ \Delta _Q \subset TQ$ are given. The kinematic and variational constraints are given by $C_K= \Delta _Q$ and $C_V= TQ \times _Q \Delta _Q$ (so that $C_V(q,v)= \Delta _Q(q)$) and the Lagrange-d'Alembert equations \eqref{LdA_NL} become
\begin{equation}\label{LdA} 
(q(t),  \dot q(t)) \in \Delta _Q(q(t)) \quad\text{and}\quad \frac{d}{dt} \frac{\partial L}{\partial \dot q}(q(t),  \dot q(t)) - \frac{\partial L}{\partial q}(q(t),  \dot q(t))  \in  \Delta _Q(q(t))^\circ.
\end{equation}
For the case of the Dirac dynamical system formulation on $P=TQ \oplus T^*Q$ studied in \S\ref{LD_Pontryagin}, the induced distribution defined in \eqref{def_tilde_CV} is $\Delta _ P (q,v,p)= \left( T_{(q, v, p)} \pi_{(P,Q)} \right) ^{-1} ( \Delta _Q(q))$ which recovers the distribution used in \eqref{DDS_NH}. 
For the case of the Lagrange-Dirac formulation on $T^*Q$ studied in the present section, the hypothesis underlying the definition \eqref{CCC_V} is clearly verified for any Lagrangian $L$ (whether $L$ is degenerate or not). We can thus define $ \mathscr{C} _V(q,p):= C_V(q,v)=\Delta _Q(q)$ and from \eqref{def_tilde_CCCV}, we have $ \Delta _{T^* Q}(q,p)= \left( T_{(q,p)} \pi _{(T^{\ast}Q,Q)}\right) ^{-1} ( \Delta _Q(q))$, which recovers the distribution used in \eqref{LD_NH}.

In conclusion, in the special case $ \mathcal{Q} =Q$, $C_K= \Delta _Q$, and $C_V= TQ \times _Q \Delta _Q$, the Dirac formulations of Theorem \ref{theorem} and Theorem \ref{theorem_LD} recover the corresponding Dirac formulations for nonholonomic mechanics \eqref{DDS_NH} and \eqref{LD_NH}.}
\end{remark}

\section{Dirac formulations induced from the thermodynamical symplectic form}\label{LD_formulation}

In this section, we first show that the equations of evolution for the thermodynamics of simple systems fall into the class that is considered in \S\ref{section_2}. Then, by applying the results of \S\ref{section_2}, we obtain the corresponding Dirac formulations for the thermodynamics of such systems, namely, a Dirac dynamical system formulation on the Pontryagin bundle and a Lagrange-Dirac formulation. We also present the associated Lagrange-d'Alembert-Pontryagin variational structures.

\subsection{Dirac formulation on the Pontryagin bundle}\label{LD_Pontryagin_Therm}

\paragraph{Constraints for the thermodynamics of simple systems.}
Let us consider a simple thermodynamic system with a Lagrangian $L=L(q,v,S): TQ \times \mathbb{R}  \rightarrow \mathbb{R}  $ and a friction force $F^{\rm fr}:TQ \times \mathbb{R}  \rightarrow T^*Q$. For simplicity, we assume that $ F^{\rm ext}=0$, see Remark \ref{Rmk_Fext}. Here $Q$ is the configuration manifold of the mechanical variables $q$ of the system, and $ \mathbb{R}  $ denotes the space of the thermodynamic variable. Let us further introduce the thermodynamic configuration manifold $\mathcal{Q} :=  Q\times \mathbb{R}$. Then, the variational constraint \eqref{CV} defines the subset
\begin{equation}\label{C_V_thermo}
C_V= \left \{(q,S, v, W,\delta q, \delta S) \in T \mathcal{Q} \times _ \mathcal{Q} T \mathcal{Q}  \;\left |\; \frac{\partial L}{\partial S}(q, v , S)\delta S= \left\langle F^{\rm fr}(q , v , S),\delta q \right\rangle\right. \right \},
\end{equation} 
where $(q,S)\in \mathcal{Q}$, $(v, W) \in T_{(q,S)} \mathcal{Q}$, and $(\delta q, \delta S) \in T_{(q,S)} \mathcal{Q}$.
Since $ \frac{\partial L}{\partial S}(q, v, S) \neq 0$ by hypothesis, see \eqref{temperature_assumption}, we obtain that $C_V$ is a submanifold of $T \mathcal{Q} \times _ \mathcal{Q} T \mathcal{Q}$ of codimension one.

The kinematic constraint $C_K$ defined from the variational constraint $C_V$ in \eqref{def_CK} is given here by
\begin{equation}\label{C_K_thermo} 
C_K=  \left \{(q,S, v, W) \in T \mathcal{Q} \;\left |\; \frac{\partial L}{\partial S}(q, v, S) W= \left\langle F^{\rm fr}(q , v , S),v \right\rangle\right. \right \},
\end{equation} 
which hence coincides with the phenomenological constraint \eqref{CK}. This justifies the terminology ``\textit{constraint of the thermodynamic type}" used for the general setting developed in \S\ref{section_2} which naturally includes the types of constraints arising in thermodynamics. 

For each fixed $(q,S, v, W) \in T\mathcal{Q} $, the annihilator of $C_V(q,S, v, W)$ is given by
\begin{equation}\label{C_V_0_thermo} 
C_V(q,S,v, W)^\circ =\left \{ (q,S, \alpha , \mathcal{T}   ) \in T^*_{(q,S)} \mathcal{Q} \;\left | \; \alpha  \frac{\partial L}{\partial S}(q, v, S)= - \mathcal{T}    F^{\rm fr}(q, v, S) \right. \right\}.
\end{equation} 
By using the expression of \eqref{C_K_thermo} and \eqref{C_V_0_thermo}, and the fact that the Lagrangian does not depend on $\dot S$, 
the system \eqref{2nd_order} yields
\[
\left\{
\begin{array}{l}
\displaystyle\vspace{0.2cm} \frac{d}{dt} \frac{\partial L}{\partial \dot q}(q, \dot q, S)- \frac{\partial L}{\partial q}(q, \dot q, S)= F^{\rm fr}(q , \dot q , S),\\
\displaystyle \frac{\partial L}{\partial S}(q, \dot q, S)\dot  S= \left\langle F^{\rm fr}(q , \dot q , S), \dot q \right\rangle.
\end{array}
\right.
\]
This recovers the equations of motion for the thermodynamics of a simple system, which is recalled in \eqref{simple_systems} with $F^{\rm ext}=0$ .

Let $\mathcal{P} =T \mathcal{Q}\oplus T ^* \mathcal{Q} $ be the Pontryagin bundle over $\mathcal{Q}$. Using the notation $x=(q,S,v,W,p, \Lambda  )$ for an element of the Pontryagin bundle $ \mathcal{P}$, the distribution induced by $C_V$ on $\mathcal{P}$, defined in \eqref{def_tilde_CV} with the help of the projection $\pi_{(\mathcal{P} ,\mathcal{Q})}: \mathcal{P} \rightarrow \mathcal{Q}$, $( q, S, v, W, p,\Lambda) \mapsto ( q, S)$ as $ \Delta _ \mathcal{P} (x):= (T _x \pi_ {(\mathcal{P},\mathcal{Q} )} ) ^{-1}(C_V(q,S,v,W))$,  reads locally
\[
\Delta _ \mathcal{P} (x):= \left\{(x, \delta x ) \in T \mathcal{P} \;\left| \; \frac{\partial L}{\partial S}(q,v,S) \delta S= \left\langle F^{\rm fr}(q,v,S), \delta q \right\rangle \right.\right  \}.
\]

\paragraph{Dirac dynamical system formulation on $ \mathcal{P} = T^* \mathcal{Q} \oplus T ^\ast \mathcal{Q} $.} 
As in \eqref{D_thermo}, we define the Dirac structure on $ \mathcal{P} $ induced from the distribution $\Delta _ \mathcal{P}$ and the presymlectic form $ \omega _{\mathcal{P} }= \pi _{(\mathcal{P} , T^* \mathcal{Q} )} ^\ast \Omega_{T^* \mathcal{Q} } $  on $ \mathcal{P} $ as
\begin{equation*}\label{D_thermo_P}
\begin{aligned}
D_{ \Delta _ \mathcal{P}}(x) :
&=\big\{ (v_{x}, \alpha_{x}) \in T_{x} \mathcal{P}  \times T^{\ast}_{x}\mathcal{P}  \; \mid \;  v_{x} \in \Delta _ \mathcal{P}(x) \; \text{and} \\ 
& \qquad \qquad \qquad \left\langle \alpha_{x},w_{x} \right\rangle =\omega_{\mathcal{P} }(x)(v_{x},w_{x}) \; \;
\mbox{for all} \; \; w_{x} \in\Delta _ \mathcal{P}(x) \big\}.
\end{aligned}
\end{equation*}

Writing locally $(x, \dot x) \in T\mathcal{P} $ and $ (x,  \zeta ) \in T^* \mathcal{P} $, where $\dot x= (\dot q,\dot S,\dot v,\dot W,\dot p, \dot \Lambda  )$, and $ \zeta =(\alpha ,  \mathcal{T}  ,  \beta , \Upsilon   , u, \Psi )$, the induced Dirac structure $D_{\Delta _ \mathcal{P}}$ on $ \mathcal{P} $ is locally described as follows: $\big((x, \dot x ),(x , \zeta  )\big) \in D_{ \Delta _ \mathcal{P}}(x)$ is equivalent to
\begin{equation}\label{local_thermo}
\left\{
\begin{array}{l} 
\displaystyle\vspace{0.2cm}(\dot p + \alpha ) \frac{\partial L}{\partial S} (q,v,S)= - (  \dot \Lambda + \mathcal{T} ) F^{\rm fr}(q,v,S),\\
\displaystyle\vspace{0.2cm}\frac{\partial L}{\partial S}(q,v,S) \dot S= \left\langle F^{\rm fr}(q,v,S), \dot q \right\rangle,\\
\displaystyle\beta =0 , \quad \Upsilon   =0, \quad u=\dot q, \quad \Psi  =\dot S.\\
\end{array}\right.
\end{equation} 
The generalized energy $ \mathcal{E} : \mathcal{P} \rightarrow \mathbb{R}  $ defined in \eqref{gen_energy} is given here by
\begin{equation}\label{GE_thermo} 
\mathcal{E} (q,S,v,W,p, \Lambda  )= \left\langle p, v \right\rangle + \Lambda  W-L(q,v,S).
\end{equation} 
Using the expression $ \mathbf{d} \mathcal{E} (q,S,v,W,p, \Lambda  )= \big(q,S,v,W,p, \Lambda  , - \frac{\partial L}{\partial q}, - \frac{\partial L}{\partial S}, p-\frac{\partial L}{\partial v},\Lambda  , v, W \big)$, the Dirac dynamical system $\big((x, \dot x), \mathbf{d} \mathcal{E} (x) \big) \in D_{\Delta _ \mathcal{P}}(x)$, for $x=( q,S,v,W,p, \Lambda ) \in \mathcal{P} $, yields
\begin{equation*}\label{implicit_thermo_phys}
\left\{
\begin{array}{l} 
\displaystyle\vspace{0.2cm} \left( \dot p- \frac{\partial L}{\partial q} (q,v,S)\right)  \frac{\partial L}{\partial S} (q,v,S)= -\left( \dot \Lambda  -\frac{\partial L}{\partial S}(q,v,S)\right) F^{\rm fr}(q,v,S),\\
\displaystyle\vspace{0.2cm}\frac{\partial L}{\partial S}(q,v,S) \dot S= \left\langle F^{\rm fr}(q,v,S), \dot q \right\rangle,\\
\displaystyle p=\frac{\partial L}{\partial v} , \quad \Lambda    =0, \quad v=\dot q, \quad W =\dot S.
\end{array}\right.
\end{equation*} 
We thus obtain the following theorem concerning the Dirac formulation of the thermodynamics of simple systems.

\begin{theorem}\label{theorem_thermo}  Consider a simple system with a Lagrangian $L=L(q,v,S): TQ \times \mathbb{R}  \rightarrow \mathbb{R}  $ and a friction force $F^{\rm fr}:TQ \times \mathbb{R}  \rightarrow T^*Q$. Then the following statements are equivalent:
\begin{itemize}
\item The curve $x(t):=\big( q(t), S(t), v(t), W(t), p(t),\Lambda  (t)\big) \in \mathcal{P}$ satisfies the equations
\begin{equation}\label{implicit_thermo_simple}
\left\{
\begin{array}{l} 
\displaystyle\vspace{0.2cm} \left( \dot p(t) - \frac{\partial L}{\partial q} (q(t) ,v (t) ,S (t) )\right)  \frac{\partial L}{\partial S} (q(t) ,v (t) ,S(t) )\\
\displaystyle\vspace{0.2cm} \qquad \qquad \qquad  = -\left( \dot \Lambda  (t) -\frac{\partial L}{\partial S}(q (t) ,v (t) ,S(t) )\right) F^{\rm fr}(q(t) ,v (t) ,S (t) ),\\
\displaystyle\vspace{0.2cm}\frac{\partial L}{\partial S}(q (t) ,v (t) ,S (t) ) \dot S (t) = \left\langle F^{\rm fr}(q (t) ,v (t) ,S (t) ), \dot q(t)  \right\rangle,\\
\displaystyle p (t) =\frac{\partial L}{\partial v} (q(t),v(t),S(t)), \quad \Lambda  (t)   =0, \quad v (t) =\dot q (t) , \quad W (t) =\dot S(t) .
\end{array}\right.
\end{equation} 
\item The curve $x(t):=( q(t), S(t), v(t), W(t), p(t), \Lambda  (t)) \in \mathcal{P}$ satisfies the {\bfi Dirac dynamical system}
\begin{equation}\label{LD_system} 
\big( (x, \dot x), \mathbf{d} \mathcal{E} (x)\big) \in D_{\Delta _ \mathcal{P}}(x).
\end{equation} 
\end{itemize}
Moreover, the system \eqref{implicit_thermo_simple} is the implicit version of the system of evolution equations
\begin{equation}\label{equations_simple_closed} 
\left\{
\begin{array}{l}
\displaystyle\vspace{0.2cm} \frac{d}{dt} \frac{\partial L}{\partial \dot q}(q (t) , \dot q (t) , S (t) )- \frac{\partial L}{\partial q}(q (t) , \dot q (t) , S(t) )= F^{\rm fr}(q(t) , \dot q (t) , S (t) ),\\
\displaystyle \frac{\partial L}{\partial S}(q (t) , \dot q (t) , S (t) )\dot  S (t) = \left\langle F^{\rm fr}(q (t) , \dot q (t) , S(t) ), \dot q (t)  \right\rangle 
\end{array}
\right.
\end{equation} 
for the thermodynamics of simple systems.
\end{theorem} 
\noindent\textbf{Proof.} The result follows from Theorem \ref{theorem} and the expressions \eqref{C_V_thermo}--\eqref{local_thermo} computed above. It also follows from the expression of the generalized energy $ \mathcal{E} : \mathcal{P} \rightarrow \mathbb{R}  $, given in \eqref{GE_thermo} and the expression of its differential $ \mathbf{d} \mathcal{E}$. $\blacksquare$

\paragraph{The Lagrange-d'Alembert-Pontryagin principle on $\mathcal{P} =T \mathcal{Q} \oplus T^* \mathcal{Q} $.} As explained in \S\ref{LD_Pontryagin}, to any Dirac dynamical system on the Pontryagin bundle with constraints of thermodynamic type, is naturally associated a variational formulation on the Pontryagin bundle, called the Lagrange-d'Alembert-Pontryagin principle. 
We now describe this variational formulation for the case of the thermodynamics of adiabatically closed simple systems.

For a curve $x(t)=\big( q(t), S(t), v(t), W(t), p(t),\Lambda  (t)\big) \in \mathcal{P} $, the Lagrange-d'Alembert-Pontryagin principle on $\mathcal{P} =T \mathcal{Q} \oplus T^* \mathcal{Q} $ given  in \eqref{LD_variational_explicit} becomes here
\begin{equation}\label{DL_variational_explicit} 
\delta \int_{t_1}^{t_2}\left[ \left\langle p, \dot q -v\right\rangle + \Lambda (\dot S-W) +L (q,v,S)\right] dt=0,
\end{equation} 
for variations $ \delta x(t)=\big( \delta q(t), \delta S(t), \delta v(t), \delta W(t), \delta p(t),\delta \Lambda  (t)\big)$ of the curve $x(t)$ that satisfy
\begin{equation}\label{DL_variational_CV}
\frac{\partial L}{\partial S}(q,v,S) \delta  S= \left\langle F^{\rm fr}(q,v,S),\delta q \right\rangle, 
\end{equation} 
and $T\pi _{( \mathcal{P} ,\mathcal{Q} )}( \delta x(t_1))=T\pi _{( \mathcal{P} ,\mathcal{Q} )}( \delta x(t_2))=0$, and where the curve $x(t)$ is subject to the phenomenological constraint
\begin{equation}\label{DL_variational_CK}
\frac{\partial L}{\partial S}(q,v,S) \dot S= \left\langle F^{\rm fr}(q,v,S),\dot q \right\rangle.
\end{equation}
From this principle, one derives the associated Lagrange-d'Alembert-Pontryagin equations given by
\begin{equation*}
\left\{
\begin{array}{l} 
\displaystyle\vspace{0.2cm} \left( \dot p- \frac{\partial L}{\partial q} (q ,v ,S )\right)  \frac{\partial L}{\partial S} (q ,v ,S )
 = -\left( \dot \Lambda -\frac{\partial L}{\partial S}(q ,v,S )\right) F^{\rm fr}(q ,v ,S ),\\
\displaystyle\vspace{0.2cm}\frac{\partial L}{\partial S}(q ,v ,S ) \dot S = \left\langle F^{\rm fr}(q ,v ,S ), \dot q  \right\rangle,\\
\displaystyle p =\frac{\partial L}{\partial v} , \quad \Lambda    =0, \quad v  =\dot q , \quad W =\dot S,
\end{array}\right.
\end{equation*} 
which are manifestly equivalent with the evolution equations for the Dirac dynamical system obtained in \eqref{implicit_thermo_simple}. 

Using the generalized energy $\mathcal{E} (q,S,v,W,p, \Lambda  )= \left\langle p, v \right\rangle +\Lambda  W-L(q,v,S)$, it follows that the variational formulation \eqref{DL_variational_explicit} can be rewritten as
\begin{equation}\label{DL_variational}
\delta \int_{t_1}^{t_2} \left[ \left\langle p, \dot q \right\rangle + \Lambda \dot S - \mathcal{E} (q,S,v,W,p, \Lambda )\right] dt=0,
\end{equation} 
for the admissible variations that satisfy \eqref{DL_variational_CV}, where the curve satisfies the nonholonomic constraint \eqref{DL_variational_CK}. Note that this principle is a Hamilton-Pontryagin version of the variational formulation developed in \cite{GBYo2016a}.

\medskip

\paragraph{The intrinsic Lagrange-d'Alembert-Pontryagin equations for thermodynamics.}
The variational formulation can be intrinsically written as follows. Consider the projection $\pi_{(\mathcal{P} ,T^{\ast}\mathcal{Q})}: \mathcal{P} \rightarrow T^*\mathcal{Q}$, $( q, S, v, W, p,\Lambda) \mapsto (q, S, p,\Lambda)$ and the one-form
$ \theta _ \mathcal{P} := \pi_{(\mathcal{P} ,T^{\ast}\mathcal{Q})}^\ast\Theta_{T^{\ast}\mathcal{Q}} \in \Omega ^1 ( \mathcal{P} )$, where $\Theta_{T^{\ast}\mathcal{Q}} $ is the canonical one-form on $ T^* \mathcal{Q} $. In coordinates, we have $ \theta _ \mathcal{P}( q, S, v, W, p,\Lambda)= p_i dq^i + \Lambda dS$.
With the help of this one-form, the variational formulation \eqref{DL_variational}--\eqref{DL_variational_CK} reads, for a curve $x (t) =( q (t) , S (t) , v (t) , W (t) , p(t) ,\Lambda(t) ) \in \mathcal{P} $,
\[
\delta \int_{t_1}^{t_2} \Big[ \!\left\langle \theta _ \mathcal{P} (x) , \dot x \right\rangle - \mathcal{E} (x)\Big] dt=0,
\]
with respect to variations $ \delta x$ such that $\delta x \in\Delta _ \mathcal{P}(x)$ and $T\pi_{(\mathcal{P} ,\mathcal{Q})}( \delta x(t_1))=T\pi _{( \mathcal{P} ,\mathcal{Q} )}( \delta x(t_2))=0$, and with the constraint $\dot x \in \widetilde{C}_V(x)$. From this principle it follows the  \textit{intrinsic Lagrange-d'Alembert-Pontryagin equations for thermodynamics} as 
\begin{equation}\label{ILDP_1} 
\mathbf{i} _ {\dot x} \omega _{ \mathcal{P} } (x)- \mathbf{d} \mathcal{E} (x) \in\Delta _ \mathcal{P}(x)^\circ, \quad \dot x \in \Delta _ \mathcal{P}(x),
\end{equation} 
where $ \omega _ \mathcal{P} = - \mathbf{d} \Theta _ \mathcal{P}$ is the presymplectic form considered earlier. In coordinates it reads $ \omega _ \mathcal{P} ( q, S, v, W, p,\Lambda) =dq ^i  \wedge dp _i +  dS \wedge d\Lambda$. Equation \eqref{ILDP_1} is the intrinsic formulation of the system \eqref{implicit_thermo_simple}.

\subsection{Lagrange-Dirac formulation}\label{LD_Cotang_Therm} 

In the above, we have presented the Dirac dynamical system formulation for the thermodynamics of simple systems, by using a Dirac structure on the Pontryagin bundle $\mathcal{P}=T \mathcal{Q} \oplus T^* \mathcal{Q} $, following the developments of \S\ref{LD_Pontryagin}. In this section we show the Lagrange-Dirac formulation for such systems by using a Dirac structure on the cotangent bundle $T^* \mathcal{Q} $, following the developments of \S\ref{LD_TsQ}.

\paragraph{Constraints for the thermodynamics of simple systems.} 
Consider a Lagrangian $L=L(q,v,S): TQ  \times \mathbb{R}  \rightarrow \mathbb{R}  $ and assume it is \textit{hyperregular with respect to the mechanical variables}, i.e., the map
\begin{equation}\label{PLT} 
\mathbb{F}L_S: TQ   \rightarrow T^* Q, \quad (q,v) \mapsto \left( q, \frac{\partial L}{\partial v}(q,v,S) \right)  
\end{equation} 
is a diffeomorphism  for each fixed $S \in \mathbb{R}  $, see \cite{GBYo2017a} for more details. Under this hypothesis, following \eqref{CCC_V}, given the variational constraint $C_V$ in \eqref{C_V_thermo}, we can define the variational constraint $\mathscr{C} _V \subset T^*\mathcal{Q} \times _\mathcal{Q}  T \mathcal{Q} $ as
\begin{equation}\label{def_CCCV} 
\mathscr{C} _V(q,S,p, \Lambda ):= C_V(q,S,v,W),
\end{equation} 
where $v$ is uniquely determined by the condition $ \frac{\partial L}{\partial v}(q,v,S)=p$. Note that by using the projection $\pi_{(T \mathcal{Q},TQ\times \mathbb{R})}: T \mathcal{Q} \to TQ \times \mathbb{R}$, $(q,S,v,W) \mapsto (q,v,S)$, we can lift the Lagrangian $L$ onto $T\mathcal{Q}$ as $\widetilde{L}=L \circ \pi_{(T \mathcal{Q},TQ\times \mathbb{R})}$. The lifted Lagrangian $\widetilde{L}:T \mathcal{Q} \rightarrow \mathbb{R}  $ is not regular  since it does not depend on $W$. 
Nevertheless, for the case of thermodynamics, the variational constraint \eqref{def_CCCV} is well-defined, because the right hand side does not depend on $W$. This is a case where the definition \eqref{CCC_V} can still be used even though the Lagrangian (here $\widetilde{L}$) is degenerate.
Explicitly, we have
\[
\mathscr{C} _V(q,S,p, \Lambda )=\{(q,S, \delta q,\delta S) \mid -T(q,p,S) \delta S= \left\langle \mathcal{F} ^{\rm fr}(q,p,S),\delta q \right\rangle \},
\]
where $T(q,p,S):= -\frac{\partial L}{\partial S}(q,v,S)$ and $\mathcal{F} ^{\rm fr}(q,p,S):= F^{\rm fr}(q,v,S)$, in which $v$ is uniquely determined from the condition $ \frac{\partial L}{\partial v}(q,v,S)=p$.

Using the notation $z=(q,S,p, \Lambda  ) \in  T ^* \mathcal{Q} $, the induced distribution on $T ^* \mathcal{Q}$, defined in \eqref{def_tilde_CCCV} with the help of the projection $ \pi _ {(T^{\ast}\mathcal{Q},\mathcal{Q})} : T ^* \mathcal{Q} \rightarrow \mathcal{Q} $ as 
$$
\Delta _{T^* \mathcal{Q} }(z):= (T _z \pi_ {(T^{\ast}\mathcal{Q},\mathcal{Q})} ) ^{-1}( \mathscr{C} _{V}(z)),  
$$
reads locally
\begin{equation}\label{loc_CVcot} 
 \Delta _{T^* \mathcal{Q} }(z):= \big\{(z, \delta z ) \in T T^*\mathcal{Q}\mid -T(q,p,S) \delta S= \left\langle  \mathcal{F} ^{\rm fr}(q,p,S), \delta q \right\rangle \big\}.
\end{equation} 

\paragraph{Lagrange-Dirac formulation on $T^* \mathcal{Q} $.} 
Given the distribution $\Delta _{T^* \mathcal{Q} }$ and the canonical symplectic form $\Omega_{T^{\ast}\mathcal{Q}}$, one can define the induced Dirac structure 
on $T^*\mathcal{Q}$ by
\begin{equation}\label{D_thermo_qSpLambda}
\begin{aligned}
D_{\Delta _{T^* \mathcal{Q} }}(z) :
&=\big\{ (v_{z}, \alpha_{z}) \in T_{z}T^*\mathcal{Q} \times T^{\ast}_{z}T^*\mathcal{Q} \; \mid \;  v_{z} \in \Delta _{T^* \mathcal{Q} }(z) \; \text{and} \\ 
& \qquad \qquad \qquad \left\langle \alpha_{z},w_{z} \right\rangle =\Omega_{T^{\ast}\mathcal{Q}}(z)(v_{z},w_{z}) \; \;
\mbox{for all} \; \; w_{z} \in \Delta _{T^* \mathcal{Q} }(z) \big\},
\end{aligned}
\end{equation}
for each $z=(q,S,p,\Lambda) \in T^{\ast}\mathcal{Q}$.
Writing locally $(z, \dot z) \in  T T^*\mathcal{Q}$ and $ (z,  \zeta ) \in T^* T^*\mathcal{Q} $, with $\dot z= (\dot q,\dot S,\dot p, \dot \Lambda  )$, and $ \zeta =(\alpha ,  \mathcal{T}  , u, \Psi )$, the condition $\big((z, \dot z ),(z, \zeta  )\big) \in D_{\Delta _{T^* \mathcal{Q} }}(z)$ is equivalent to
\[
\left\{
\begin{array}{l} 
\displaystyle\vspace{0.2cm}(\dot p + \alpha ) T(q,p,S)=  ( \dot \Lambda+ \mathcal{T}    )\mathcal{F} ^{\rm fr}(q,p,S),\\
\displaystyle\vspace{0.2cm}T(q,p,S) \dot S=- \left\langle \mathcal{F} ^{\rm fr}(q,p,S), \dot q \right\rangle,\\
\displaystyle  u=\dot q, \quad \Psi  =\dot S.
\end{array}\right.
\]
From this local expression, it follows that
$$
\big( (q,S,p, \Lambda , \dot q, \dot S,\dot p, \dot\Lambda ), \mathbf{d} _D\widetilde{L}(q,S,v,W)\big) \in D_{\Delta _{T^* \mathcal{Q} }}(q,S,p, \Lambda )
$$
if and only if
\[
\left\{
\begin{array}{l} 
\displaystyle\vspace{0.2cm} \left(  \dot p- \frac{\partial L}{\partial q}(q,v,S) \right)  T(q,p,S)= \left( \dot \Lambda -\frac{\partial L}{\partial S}(q,v,S) \right)  \mathcal{F} ^{\rm fr}(q,p,S),\\
\displaystyle\vspace{0.2cm}T(q,S,p) \dot S= -  \left\langle \mathcal{F} ^{\rm fr}(q,p,S), \dot q \right\rangle,\\
\displaystyle  v=\dot q, \quad W  =\dot S, \quad p=\frac{\partial L}{\partial v}(q,v,S), \quad \Lambda =0,\\
\end{array}\right.
\]
where the last two equalities come from the fact that $(q,S,p, \Lambda , \dot q, \dot S,\dot p, \dot\Lambda )$ and $\mathbf{d} _D\widetilde{L}(q,S,v,W)$ both belong to fibers at the point $(q,S,p, \Lambda ) \in T^{\ast}\mathcal{Q}$. Recall that
\[
\mathbf{d} _D\widetilde{L}(q,S,v,W)= \Big( \gamma _ \mathcal{Q} \circ \mathbf{d} \widetilde{L}\Big )(q,S,v,W)=  \left( q, S,\frac{\partial L}{\partial v},0, - \frac{\partial L}{\partial q},- \frac{\partial L}{\partial S}, v, W \right).
\]
One thus obtains the following Lagrange-Dirac formulation for the thermodynamic of simple systems.

\begin{theorem}\label{theorem_thermo_DL}  Consider a simple system with a Lagrangian $L=L(q,v,S): TQ \times \mathbb{R}  \rightarrow \mathbb{R}  $ and a friction force $F^{\rm fr}:TQ \times \mathbb{R}  \rightarrow T^*Q$. Assume that $L$ is hyperregular with respect to the mechanical variables $(q,v)$ and define $T(q,p,S)$ and $ \mathcal{F} ^{\rm fr}(q,p,S)$ as before. Then the following statements are equivalent:
\begin{itemize}
\item The curve $\big( q(t), S(t), v(t), W(t), p(t),\Lambda  (t)\big) \in \mathcal{P}$ satisfies the equations
\begin{equation}\label{implicit_thermo_simple_DL}
\left\{
\begin{array}{l} 
\displaystyle\vspace{0.2cm} \left( \dot p(t) - \frac{\partial L}{\partial q} (q(t) ,v (t) ,S (t) )\right)  T (q(t) ,p (t) ,S(t) )\\
\displaystyle\vspace{0.2cm} \qquad \qquad \qquad  = \left( \dot \Lambda  (t) -\frac{\partial L}{\partial S}(q (t) ,v (t) ,S(t) )\right) \mathcal{F} ^{\rm fr}(q(t), p(t) ,S (t) ),\\
\displaystyle\vspace{0.2cm} T (q(t) ,p (t) ,S(t) )\dot S (t) =- \left\langle \mathcal{F} ^{\rm fr}(q (t), p(t), S(t) ), \dot q(t)  \right\rangle,\\
\displaystyle p (t) =\frac{\partial L}{\partial v}(q(t),v(t),S(t)) , \quad \Lambda  (t)   =0, \quad v (t) =\dot q (t) , \quad W (t) =\dot S(t) .
\end{array}\right.
\end{equation} 
\item The curve $(q(t), S(t), v(t), W(t), p(t), \Lambda  (t)) \in \mathcal{P}$ satisfies the {\bfi Lagrange-Dirac system}
\[
\big( (q ,S ,p , \Lambda , \dot q , \dot S  ,\dot p , \dot\Lambda   ), \mathbf{d} _D\widetilde{L}(q ,S   ,v   ,W  )\big) \in D_{ \Delta _{T^* \mathcal{Q} }}(q  ,S  ,p , \Lambda  ).
\]
\end{itemize}
Moreover, the system \eqref{implicit_thermo_simple_DL} is an implicit version of the system of evolution equations \eqref{equations_simple_closed} for the thermodynamics of simple systems.
\end{theorem}

\section{Dirac formulations induced from the mechanical symplectic form}\label{section_4}

In this section, we show that Dirac formulations of nonequilibrium thermodynamics of simple systems can be constructed by using the canonical symplectic form $ \Omega _{T^*Q}$ on the cotangent bundle of the mechanical configuration manifold $Q$. In this case, we can develop a Hamilton-Dirac description.

\subsection{Dirac formulation on $ M= TQ \oplus  _ \mathcal{Q} T ^\ast Q$}\label{DF_4}

We consider the vector bundle on $ \mathcal{Q} $ given by
\[
\pi _{(M, \mathcal{Q} )}:  M= TQ \oplus  _ \mathcal{Q} T ^\ast Q \rightarrow \mathcal{Q},
\]
whose vector fiber at $(q,S) \in \mathcal{Q} $ is $T_qQ \times T^*_qQ$. An element in this fiber is denoted as $(q,S,v,p)$. Note that we can write $M$ as the pull-back of the Pontryagin bundle of $Q$, i.e., $M= \pi ^\ast _{ (\mathcal{Q} , Q)} (TQ \oplus T^*Q)$, where $ \pi _{( \mathcal{Q} , Q)}: \mathcal{Q} \rightarrow Q$ is the projection $ \pi _{( \mathcal{Q} , Q)}(q,S)=q$.

\paragraph{Constraints for the thermodynamics of simple systems.} In \S\ref{LD_Pontryagin_Therm} we have considered the distribution $ \Delta _ \mathcal{P} $ on $ \mathcal{P} = T \mathcal{Q} \oplus T^* \mathcal{Q} $ induced by the variational constraint $C_V$ in \eqref{C_V_thermo} associated to \eqref{CV}. We observe that the variational constraint $C_V(q,S,v,W) \subset T \mathcal{Q} $ does not depend on $W$, so it can be used to yield a distribution on $M$ as
\[
\Delta _M(q,S,v,p):= \left( T _{(q,S,v,p)} \pi _{ (M, \mathcal{Q}) } \right) ^{-1} \left( C_V(q,S,v,W) \right) . 
\]
It is locally given as
\[
\Delta _ M (q,S,v,p):= \left\{(q,S,v,p, \delta q,\delta S,\delta v,\delta p ) \in T M \;\left| \; \frac{\partial L}{\partial S}(q,v,S) \delta S= \left\langle F^{\rm fr}(q,v,S), \delta q \right\rangle \right.\right  \}.
\]

\paragraph{Dirac formulation on $ M = TQ\oplus  _\mathcal{Q}  T ^\ast Q$.} Consider the projection 
\[
\pi _{(M, T^* Q )}: M \rightarrow T^* Q ,\quad \pi _{(M, T^* Q)}(q,S,v,p):= (q,p)
\]
and the presymplectic form $ \omega _{M }= \pi _{(M , T^*Q )} ^\ast \Omega_{T^* Q} $  on $ M $ induced from the canonical symplectic form of $T^*Q$.
Denoting by $m=(q,S,v,p)$ an arbitrary element in $M$, we define the Dirac structure on $ M $ induced from the distribution $\Delta _ M$ and the presymlectic form $ \omega _M$  as
\begin{equation*}\label{D_thermo_P}
\begin{aligned}
D_{ \Delta _ M}(m) :
&=\big\{ (v_{m}, \zeta _{m}) \in T_{m} M  \times T^{\ast}_{m}M  \; \mid \;  v_{m} \in \Delta _ M(m) \; \text{and} \\ 
& \qquad \qquad \qquad \left\langle \zeta_{m},w_{m} \right\rangle =\omega_{M }(m)(v_{m},w_{m}) \; \;
\mbox{for all} \; \; w_{m} \in\Delta _ M(m) \big\}.
\end{aligned}
\end{equation*}

Writing locally $(m, \dot m) \in TM $ and $ (m,  \zeta ) \in T^* M $, where $\dot m= (\dot q,\dot S,\dot v, \dot p  )$, and $ \zeta =(\alpha ,  \mathcal{T}  ,  \beta    , u )$, the induced Dirac structure $D_{\Delta _ M}$ on $ M $ is locally described as follows: for each $m=( q, S, v,p)$, the condition  $\big((m, \dot m ),(m , \zeta  )\big) \in D_{ \Delta _ M}(m)$ is equivalent to
\[
\left\{
\begin{array}{l} 
\displaystyle\vspace{0.2cm}(\dot p + \alpha ) \frac{\partial L}{\partial S} (q,v,S)= -  \mathcal{T}  F^{\rm fr}(q,v,S),\\
\displaystyle\vspace{0.2cm}\frac{\partial L}{\partial S}(q,v,S) \dot S= \left\langle F^{\rm fr}(q,v,S), \dot q \right\rangle,\\
\displaystyle\beta =0 , \quad u=\dot q.\\
\end{array}\right.
\]
The generalized energy $E : M \rightarrow \mathbb{R}  $ is defined here by
\[
E(q,S,v,p  )= \left\langle p, v \right\rangle -L(q,v,S).
\]
Using the expression $ \mathbf{d}E (q,S,v,p  )= \Big(q,S,v,p  , - \frac{\partial L}{\partial q}, - \frac{\partial L}{\partial S}, p-\frac{\partial L}{\partial v}, v\Big)$, the Dirac dynamical system $\big((m, \dot m), \mathbf{d} E(m) \big) \in D_{\Delta _ M}(m)$, for $m=( q,S,v,p ) \in M $, yields
\[
\left\{
\begin{array}{l} 
\displaystyle\vspace{0.2cm} \left( \dot p- \frac{\partial L}{\partial q} (q,v,S)\right)  \frac{\partial L}{\partial S} (q,v,S)= \frac{\partial L}{\partial S}(q,v,S) F^{\rm fr}(q,v,S),\\
\displaystyle\vspace{0.2cm}\frac{\partial L}{\partial S}(q,v,S) \dot S= \left\langle F^{\rm fr}(q,v,S), \dot q \right\rangle,\\
\displaystyle p=\frac{\partial L}{\partial v} , \quad v=\dot q.
\end{array}\right.
\]
We thus obtain the following theorem concerning the Dirac formulation of the thermodynamics of simple systems on $ M$.

\begin{theorem}\label{theorem_thermo}  Consider a simple system with a Lagrangian $L=L(q,v,S): TQ \times \mathbb{R}  \rightarrow \mathbb{R}  $ and a friction force $F^{\rm fr}:TQ \times \mathbb{R}  \rightarrow T^*Q$. Then the following statements are equivalent:
\begin{itemize}
\item The curve $m(t):=( q(t), S(t), v(t) , p(t)) \in M$ satisfies the equations
\begin{equation}\label{implicit_thermo_simple_4}
\left\{
\begin{array}{l} 
\displaystyle\vspace{0.2cm}   \dot p(t) - \frac{\partial L}{\partial q} (q(t) ,v (t) ,S (t) ) =  F^{\rm fr}(q(t) ,v (t) ,S (t) ),\\
\displaystyle\vspace{0.2cm}\frac{\partial L}{\partial S}(q (t) ,v (t) ,S (t) ) \dot S (t) = \left\langle F^{\rm fr}(q (t) ,v (t) ,S (t) ), \dot q(t)  \right\rangle,\\
\displaystyle p (t) =\frac{\partial L}{\partial v} (q(t),v(t),S(t)), \quad v (t) =\dot q (t) .
\end{array}\right.
\end{equation} 
\item The curve $m(t):=( q(t), S(t), v(t),  p(t)) \in M$ satisfies the {\bfi Dirac dynamical system}
\begin{equation}\label{LD_system_4} 
\big( (m, \dot m), \mathbf{d} E (m)\big) \in D_{\Delta _ M}(m).
\end{equation} 
\end{itemize}
Moreover, the system \eqref{implicit_thermo_simple_4} is the implicit version of the system of evolution equations
\[
\left\{
\begin{array}{l}
\displaystyle\vspace{0.2cm} \frac{d}{dt} \frac{\partial L}{\partial \dot q}(q (t) , \dot q (t) , S (t) )- \frac{\partial L}{\partial q}(q (t) , \dot q (t) , S(t) )= F^{\rm fr}(q(t) , \dot q (t) , S (t) ),\\
\displaystyle \frac{\partial L}{\partial S}(q (t) , \dot q (t) , S (t) )\dot  S (t) = \left\langle F^{\rm fr}(q (t) , \dot q (t) , S(t) ), \dot q (t)  \right\rangle 
\end{array}
\right.
\]
for the thermodynamics of simple systems.
\end{theorem}

\paragraph{Lagrange-d'Alembert-Pontryagin principle on $M$.} Let us now describe the variational formulation associated to the above Dirac dynamical system.
Given a curve $m(t)=( q(t), S(t), v(t),   p(t) ) \in M $, we consider the Lagrange-d'Alembert-Pontryagin principle on $M $ given by
\begin{equation}\label{DL_variational_explicit_4} 
\delta \int_{t_1}^{t_2}\Big[ \left\langle p, \dot q -v\right\rangle +L (q,v,S)\Big] dt=0,
\end{equation} 
for variations $ \delta m(t)=( \delta q(t), \delta S(t), \delta v(t), \delta p(t) (t))$ of the curve $m(t)$ that satisfy
\begin{equation}\label{DL_variational_CV_4}
\frac{\partial L}{\partial S}(q,v,S) \delta  S= \left\langle F^{\rm fr}(q,v,S),\delta q \right\rangle, 
\end{equation} 
and $T\pi _{( M , Q )}( \delta m(t_1))=T\pi _{( M ,Q )}( \delta m(t_2))=0$, and where the curve $m(t)$ is subject to the phenomenological constraint
\begin{equation}\label{DL_variational_CK_4}
\frac{\partial L}{\partial S}(q,v,S) \dot S= \left\langle F^{\rm fr}(q,v,S),\dot q \right\rangle.
\end{equation}
It is easily verified that this principle yields the implicit system \eqref{implicit_thermo_simple_4} on $M$.

\medskip

\paragraph{Intrinsic Lagrange-d'Alembert-Pontryagin equations on $M$.}
The above variational formulation can be intrinsically written as follows. Consider the projection $\pi_{(M ,T^{\ast}Q)}: M \rightarrow T^*Q$, $( q, S, v, p) \mapsto (q, p)$ and the one-form
$ \theta _ M := \pi_{(M ,T^{\ast}Q)}^\ast\Theta_{T^{\ast}Q} \in \Omega ^1 ( M )$, where $\Theta_{T^{\ast}Q} $ is the canonical one-form on $ T^*Q$. In coordinates, we have $ \theta _ M( q, S, v,p)= p_i dq^i $.
With the help of this one-form, the variational formulation \eqref{DL_variational_explicit_4}--\eqref{DL_variational_CK_4} reads, for a curve $m (t) =( q (t) , S (t) , v (t) , p (t) ) \in M $,
\[
\delta \int_{t_1}^{t_2} \Big[ \!\left\langle \theta _ M (m) , \dot m \right\rangle - E(m)\Big] dt=0,
\]
with respect to variations $ \delta m$ of the curve such that $\delta m \in\Delta _ M(m)$ and $T\pi_{(M ,Q)}( \delta m(t_1))=T\pi _{( M ,Q)}( \delta m(t_2))=0$, and with the constraint $\dot m \in \Delta _M(m)$. From this principle it follows the  \textit{intrinsic Lagrange-d'Alembert-Pontryagin equations for thermodynamics} on $M$ as 
\begin{equation}\label{ILDP_1_4} 
\mathbf{i} _ {\dot m} \omega _{ M } (m)- \mathbf{d}E (m) \in\Delta _ M(m)^\circ, \quad \dot m \in \Delta _ M(m),
\end{equation} 
where $ \omega _ M = - \mathbf{d} \theta _ M$ is the presymplectic form considered above. Equation \eqref{ILDP_1_4} is the intrinsic formulation of the system \eqref{implicit_thermo_simple_4}.

\subsection{Lagrange-Dirac formulation on $N=T^*Q \times \mathbb{R}  $}\label{LD_4} 

We describe here a Lagrange-Dirac formulation on $N=T^*Q \times \mathbb{R}  $ associated to the Dirac structure on $N$ induced from the variational constraint $C_V \subset T \mathcal{Q}\times _ \mathcal{Q} T \mathcal{Q}$ and the canonical symplectic form on $T^*Q$.

\paragraph{Constraints.} Consider a Lagrangian $L=L(q,v,S): TQ  \times \mathbb{R}  \rightarrow \mathbb{R}  $ of a simple thermodynamic system and assume it is \textit{hyperregular with respect to the mechanical variables} $(q,v)$, see \eqref{PLT}. Recall that in this case we can define from $C_V$ the constraint $\mathscr{C} _V(q,S,p,\Lambda)\in T_{(q,S)} \mathcal{Q}$, see \eqref{def_CCCV}.
Since $\mathscr{C} _V$ does not depend on $W$ it induces the following distribution on $N$:
\[
\Delta _ N(q,S,p):= \left( T_{(q,S,p)} \pi _{(N, \mathcal{Q})} \right) ^{-1} \big( \mathscr{C} _V(q,S,p,\Lambda) \big),
\]
locally given as
\[
\Delta _ N (q,S,p)= \left\{(q,S,p, \delta q,\delta S,\delta p ) \in T N \;\left| \; -T(q,S,p) \delta S= \left\langle  \mathcal{F} ^{\rm fr}(q,S,p), \delta q \right\rangle \right.\right  \}.
\]
We recall the definition $T(q,S,p)= - \frac{\partial L}{\partial S}(q,v,S)$, in which $v$ is uniquely defined from the condition $ \frac{\partial L}{\partial v}(q,v,S)=p$. 
Associated to this distribution and to the presymplectic form $\omega _N= \pi _{(N, T^*Q)} ^\ast \Omega _{T^*Q}$, is the Dirac structure on $N$ given by, for each $n=(q,S,p) \in N$,
\begin{equation*}\label{D_thermo_N}
\begin{aligned}
D_{ \Delta _ N}(n) :
&=\big\{ (v_{n}, \zeta _{n}) \in T_{n} N  \times T^{\ast}_{n}N  \; \mid \;  v_{n} \in \Delta _ N(n) \; \text{and} \\ 
& \qquad \qquad \qquad \left\langle \zeta_{n},w_{n} \right\rangle =\omega_{N }(n)(v_{n},w_{n}) \; \;
\mbox{for all} \; \; w_{n} \in\Delta _ N(n) \big\}.
\end{aligned}
\end{equation*}
Writing locally $(n, \dot n) \in TN $ and $ (n,  \zeta ) \in T^* N $, where $\dot n= (\dot q,\dot S, \dot p  )$, and $ \zeta =(\alpha ,  \mathcal{T}  , u )$, the induced Dirac structure $D_{\Delta _ N}$ on $ N $ is locally described as follows: for each $n=( q, S, p)$, the condition  $\big((n, \dot n ),(n , \zeta  )\big) \in D_{ \Delta _ N}(n)$ is equivalent to
\begin{equation}\label{local_thermo_LD4}
\left\{
\begin{array}{l} 
\displaystyle\vspace{0.2cm}(\dot p+ \alpha ) T (q,S,p)=  \mathcal{T}   \mathcal{F} ^{\rm fr}(q,S,p),\\
\displaystyle\vspace{0.2cm}T(q,S,p) \dot S= -\left\langle\mathcal{F} ^{\rm fr}(q,S,p), \dot q \right\rangle,\\
\displaystyle u=\dot q.\\
\end{array}\right.
\end{equation}

\paragraph{Lagrange-Dirac formulation on $N$.} Recall that the Dirac differential for a Lagrangian $L:TQ \rightarrow \mathbb{R}  $ is defined by using the canonical diffeomorphism $ \gamma _Q: T^*TQ \rightarrow T^*T^*Q$, locally given by $\gamma _Q(q, v, \alpha , p)=(q,p,-  \alpha ,v)$. In our case, we extend $ \gamma _Q$ to the diffeomorphism
\[
\widehat{ \gamma }_Q: T^*(TQ \times \mathbb{R}  ) \rightarrow T^*(T^*Q \times \mathbb{R}  ), \quad \widehat{ \gamma }_Q(q,S,v, \alpha  ,\Lambda, p  ):=(q,S,p,-\alpha  , -\Lambda, v),
\]
which is operating on $T^* \mathbb{R}  $ as $( S, \Lambda ) \mapsto (S, - \Lambda )$. Note that $\widehat{ \gamma }_Q$ is a symplectic diffeomorphism with respect to the canonical symplectic structures $\Omega_{T^\ast(TQ \times \mathbb{R}  )} =-\mathbf{d}\Theta_{T^\ast(TQ \times \mathbb{R} ) }$ and $\Omega_{T^\ast(T^\ast Q \times \mathbb{R}  )}=+\mathbf{d}\Theta_{T^\ast(T^\ast Q \times \mathbb{R} ) }$, where 
\begin{align*} 
\Theta_{T^\ast(TQ \times \mathbb{R} ) }(q,S, v,\alpha, \Lambda , p)&= \alpha dq + p d v + \Lambda dS,\\
\Theta_{T^\ast(T^\ast Q \times \mathbb{R} )}(q,S,p, \alpha, \mathcal{T}, u)&= \alpha dq + u dp + \mathcal{T} dS.
\end{align*}

The associated Dirac differential of $L$ is
\[
\widehat{\mathbf{d}}_DL(q,S,v):=  \left( \widehat{\gamma }_ Q \circ \mathbf{d} L\right)(  q,S,v) =  \left( q, S,\frac{\partial L}{\partial v},- \frac{\partial L}{\partial q},-\frac{\partial L}{\partial S}, v\right).
\]
From this definition and the expression \eqref{local_thermo_LD4} of the Dirac structure, we have
\[
\big( (q,S,p, \dot q, \dot S, \dot p), \widehat{\mathbf{d}} _D L(q,S,v) \big) \in D_{ \Delta _N}(q,S,p)
\]
if and only if
\[
\left\{
\begin{array}{l} 
\displaystyle\vspace{0.2cm} \left(\dot p - \frac{\partial L}{\partial q}(q,v,S) \right)  T(q,p,S)= -\frac{\partial L}{\partial S}(q,v,S)  \mathcal{F} ^{\rm fr}(q,p,S),\\
\displaystyle\vspace{0.2cm}T(q,S,p) \dot S= -\left\langle \mathcal{F} ^{\rm fr}(q,p,S), \dot q \right\rangle,\\
\displaystyle  v=\dot q,  \quad p=\frac{\partial L}{\partial v}(q,v,S),\\
\end{array}\right.
\]
where the last equality comes from the fact that $(q,S,p , \dot q, \dot S,\dot p )$ and $\widehat{\mathbf{d}} _DL(q,S,v)$ both belong to the fibers at $(q,S,p ) \in T^{\ast}Q \times \mathbb{R}  $. One thus obtains the following Lagrange-Dirac formulation for the thermodynamic of simple systems, based on the canonical symplecic form $ \Omega _{T^*Q}$.

\begin{theorem}\label{theorem_thermo_DL_4}  Consider a simple system with a Lagrangian $L=L(q,v,S): TQ \times \mathbb{R}  \rightarrow \mathbb{R}  $ and a friction force $F^{\rm fr}:TQ \times \mathbb{R}  \rightarrow T^*Q$. Assume that $L$ is hyperregular with respect to the mechanical variables $(q,v)$ and define $T(q,p,S)$ and $ \mathcal{F} ^{\rm fr}(q,p,S)$ as before. Then the following statements are equivalent:
\begin{itemize}
\item The curve $( q(t), S(t), v(t), p(t)) \in M$ satisfies the equations
\begin{equation}\label{implicit_thermo_simple_DL_4}
\left\{
\begin{array}{l} 
\displaystyle\vspace{0.2cm} \left( \dot p(t) - \frac{\partial L}{\partial q} (q(t) ,v (t) ,S (t) )\right)  T (q(t) ,p (t) ,S(t) )\\
\displaystyle\vspace{0.2cm} \qquad \qquad \qquad  =  -\frac{\partial L}{\partial S}(q (t) ,v (t) ,S(t) )\mathcal{F} ^{\rm fr}(q(t), p(t) ,S (t) ),\\
\displaystyle\vspace{0.2cm} T(q(t),v(t),S(t))\dot S (t) = -\left\langle \mathcal{F} ^{\rm fr}(q (t), p(t), S(t) ), \dot q(t)  \right\rangle,\\
\displaystyle v (t) =\dot q (t) , \quad p (t) =\frac{\partial L}{\partial v}(q(t),v(t),S(t)) .
\end{array}\right.
\end{equation} 
\item The curve $(q(t), S(t), v(t), p(t)) \in M$ satisfies the {\bfi Lagrange-Dirac system}
\[
\big( (q ,S ,p ,  \dot q , \dot S  ,\dot p   ),\widehat{\mathbf{d}} _DL (q ,S   ,v  )\big) \in D_{ \Delta _{N }}(q  ,S  ,p  ).
\]
\end{itemize}
Moreover, the system \eqref{implicit_thermo_simple_DL_4} is an implicit version of the system of evolution equations \eqref{equations_simple_closed} for the thermodynamics of simple systems.
\end{theorem}

\subsection{Hamilton-Dirac formulation on $N=T^*Q \times \mathbb{R}$}\label{HD_4}

We assume that $L:TQ \times \mathbb{R} \rightarrow \mathbb{R}  $ is
\textit{hyperregular with respect to the mechanical variables}, see \eqref{PLT}.
Under this assumption, we can define the \textit{Hamiltonian function} $H:N=T^* Q \times \mathbb{R}  \rightarrow \mathbb{R}$ by
\begin{equation}\label{def_H} 
H(q,p,S)=\left\langle p, \dot q \right\rangle -L(q, \dot q, S),
\end{equation} 
where $\dot q$ is uniquely determined from $(q,p,S)$ by the condition $\frac{\partial L}{\partial \dot q}(q, \dot q, S)=p$. 

We shall use the same distribution and the same Dirac structure as in \S\ref{LD_4} before. 
Note that in \eqref{local_thermo_LD4} we can directly write the constraint in terms of the Hamiltonian as $T(q,S,p)= \frac{\partial H}{\partial S}(q,S,p)$. 
From \eqref{local_thermo_LD4}, it follows that the Hamilton-Dirac system 
\[
\big( (q,S,p, \dot q, \dot S, \dot p), \mathbf{d}H(q,S,p) \big) \in D_{ \Delta _N}(q,S,p)
\]
is equivalent to
\begin{equation}\label{local_thermo_HD4}
\left\{
\begin{array}{l} 
\displaystyle\vspace{0.2cm} \left(  \dot p+  \frac{\partial H}{\partial q}(q,S,p) \right)   \frac{\partial H}{\partial S}(q,S,p) = \frac{\partial H}{\partial S}(q,S,p)  \mathcal{F} ^{\rm fr}(q,S,p),\\
\displaystyle\vspace{0.2cm} \frac{\partial H}{\partial p} =\dot q, \qquad \frac{\partial H}{\partial S}(q,S,p)  \dot S= -\left\langle\mathcal{F} ^{\rm fr}(q,S,p), \dot q \right\rangle.
\end{array}\right.
\end{equation}
We get the following theorem.

\begin{theorem} Consider a simple system with a Lagrangian $L=L(q,v,S): TQ \times \mathbb{R}  \rightarrow \mathbb{R}  $ and a friction force $F^{\rm fr}:TQ \times \mathbb{R}  \rightarrow T^*Q$. Assume that the Lagrangian is hyperregular with respect to the mechanical variables, consider the associated Hamiltonian $H: T^*Q \times \mathbb{R}  \rightarrow \mathbb{R}$ and define $ \mathcal{F} ^{\rm fr}(q,p,S)$ as before.
Then the following statements are equivalent:
\begin{itemize}
\item The curve $( q(t), S(t),   p(t)) \in N$ satisfies the equations
\begin{equation}\label{implicit_thermo_simple_HD_4}
\left\{
\begin{array}{l} 
\displaystyle\vspace{0.2cm} \left( \dot p(t) +\frac{\partial H}{\partial q} (q(t) ,p(t) ,S (t) )\right)  \frac{\partial H}{\partial S} (q(t) ,p(t) ,S (t) )\\
\displaystyle\vspace{0.2cm} \qquad \qquad \qquad  = \frac{\partial H}{\partial S} (q(t) ,p(t) ,S (t) )\mathcal{F} ^{\rm fr}(q(t), p(t) ,S (t) ),\\
\displaystyle\vspace{0.2cm}-\frac{\partial H}{\partial S} (q(t) ,p(t) ,S (t) )\dot S (t) = \left\langle \mathcal{F} ^{\rm fr}(q (t), p(t), S(t) ), \dot q(t)  \right\rangle,\\
\displaystyle \frac{\partial H}{\partial p}(q(t),p(t),S(t))=\dot q(t).
\end{array}\right.
\end{equation} 
\item The curve $(q(t), S(t),   p(t)) \in N$ satisfies the {\bfi Hamilton-Dirac system}
\begin{equation}\label{HD_system_HD}
\begin{aligned}
&\big( (q,S,p, \dot q, \dot S, \dot p), \mathbf{d}H(q,S,p) \big) \in D_{ \Delta _N}(q,S,p).
\end{aligned}
\end{equation} 
\end{itemize}
Moreover the system \eqref{implicit_thermo_simple_HD_4}, equivalently written as
\begin{equation}\label{Hamiltonian_side}
\left\{
\begin{array}{l} 
\displaystyle\vspace{0.2cm} \dot p(t) =- \frac{\partial H}{\partial q} (q(t) ,p(t) ,S (t) )+\mathcal{F} ^{\rm fr}(q(t), p(t) ,S (t) ),\\
\displaystyle\vspace{0.2cm} \dot q(t)=\frac{\partial H}{\partial p}(q(t),p(t),S(t)),\\
\displaystyle\frac{\partial H}{\partial S} (q(t) ,p(t) ,S (t) )\dot S (t) = -\left\langle \mathcal{F} ^{\rm fr}(q (t), p(t), S(t) ), \dot q(t)  \right\rangle\\
\end{array}\right.
\end{equation} 
is the Hamiltonian description of the system of evolution equations \eqref{equations_simple_closed} for the thermodynamics of simple systems.
\end{theorem} 

\paragraph{Hamilton-d'Alembert principle on $N$.} To the Hamilton-Dirac formulation is naturally associated a variational structure. In our case, the variational formulation on $N$ is
\begin{equation}\label{HD_variational_II} 
\delta \int_{t_1}^{t_2} \Big[\! \left\langle p, \dot q \right\rangle - H (q,S,p )\Big] dt=0
\end{equation} 
for all variations $(\delta q(t), \delta S(t), \delta p (t))$ for the curve $(q(t),S(t), p(t)) \in N$ that satisfy
\begin{equation}\label{HD_variational_CV} 
-\frac{\partial H}{\partial S}(q,p,S) \delta  S= \left\langle \mathcal{F} ^{\rm fr}(q,p,S),\delta q \right\rangle 
\end{equation}
with $\delta q(t_1 )= \delta q(t _2 )=0$, and the curve is subject to the phenomenological constraint
\begin{equation}\label{HD_variational_CK} 
-\frac{\partial H}{\partial S}(q,p,S) \dot S= \left\langle\mathcal{F} ^{\rm fr}(q,p,S),\dot q \right\rangle .
\end{equation}
We refer to the principle \eqref{HD_variational_II}--\eqref{HD_variational_CK} as the \textit{Hamilton-d'Alembert principle}. From this principle one immediately obtains the system \eqref{Hamiltonian_side}.

\paragraph{Intrinsic form of the Hamilton-d'Alembert principle on $N$.}
The variational formulation can be intrinsically written as follows. 
Let us define $ \theta _ N :=   \pi_ {(N, T^\ast  Q)}\Theta _{T^*Q }\in \Omega ^1 ( N )$, where we recall that $\pi_ {(N, T^\ast  Q)}: N \rightarrow T^*  Q $, and $\Theta_{T^* Q } $ is the canonical one-form on $ T^* Q$.

The variational formulation is intrinsically described in terms of a curve $n(t) \in  N $ as
\[
\delta \int_{t_1}^{t_2}\Big[\! \left\langle \theta _ N (n) , \dot n\right\rangle - H(n)\Big] dt=0,
\]
for all variations $ \delta n$ such that $\delta n\in \Delta _N(n(t)) $ and $T\pi_{( N , Q )}( \delta n(t_1))=T\pi _{(\mathcal{M}  ,\mathcal{Q} )}( \delta n(t_2))=0$, and with the constraint $ \dot n(t) \in \Delta _N(n(t))$.
It yields the following \textit{intrinsic Hamilton-d'Alembert equation on $N$}
\begin{equation}\label{intrinsic_H_1} 
\mathbf{i} _ { \dot n} \omega _{N}(n) - \mathbf{d} H(n) \in \Delta _N(n)^\circ, \quad \dot n \in \Delta _N(n),
\end{equation} 
where $ \omega _N=- \mathbf{d} \theta _ N$. Equation \eqref{intrinsic_H_1} is  the intrinsic expression of the system \eqref{Hamiltonian_side}. 

\begin{remark}[Recovering the canonical formulation of classical mechanics]{\rm When $H$ does not depend on $S$ and the friction force is absent, it goes without saying that \eqref{intrinsic_H_1} reduces to the canonical symplectic formulation of Hamiltonian mechanics, namely,
\[
\mathbf{i} _ {( \dot q, \dot p)}\Omega _{T^*Q}(q,p) = \mathbf{d} H(q,p).
\]
}
\end{remark}

\paragraph{Relation between the Dirac formulations on $N$ and $M$.} The Hamilton-Dirac formulation on $ N$ presented in this section, namely
\begin{equation}\label{HD_M} 
\big((q,S,p, \dot q, \dot S, \dot p), \mathbf{d}H(q,S,p)\big) \in D_{\Delta _N}(q,S,p)
\end{equation}
is related, via the partial Legendre transform, to the Dirac formulation on $M$ presented in \S\ref{DF_4} given by
\begin{equation}\label{LD_P} 
\big((q,S,v,p, \dot q, \dot S, \dot v, \dot p), \mathbf{d} E (q,S,v,p)\big) \in D_{\Delta _M}(q,S,v,p).
\end{equation}
Indeed, the partial Legendre transform induces the fiber preserving immersion
\[
j_L: N \hookrightarrow M,\quad
j_L(q,S,p ):= \big(q,S, v(q,p,S), p\big),
\]
over $ \mathcal{Q} $, where the function $v=v(q,p,S)$ is defined by such that the condition $ \mathbb{F}  L_S(q,v)=(q,p)$ holds.

One notices the following relations:
\[
E \circ j_L= H ,\quad j_L ^\ast \theta _M =\theta _ N ,\quad  j_L ^\ast \omega  _ M = \omega  _ N,
\]
\[
T_{(q,S,p)}j_L\big(\Delta _N(q,S,p)\big) = \Delta _M( j_L(q,S,p))\cap \operatorname{Im}\big(T_{(q,S,p)} j_L\big).
\]
By using the explicit forms \eqref{implicit_thermo_simple_4}, resp., \eqref{Hamiltonian_side} of the systems \eqref{LD_P}, resp., \eqref{HD_M}  we notice that if the curve $n(t) \in N $ is a solution of the Hamilton-Dirac system \eqref{HD_M}, then the curve $m(t) \in M$ defined by $m(t):= j_L(n(t))$ is a solution of the Dirac dynamical system \eqref{LD_P}. Conversely, if the curve $m(t) \in M $ is a solution of the Dirac dynamical system \eqref{LD_P}, then the curve is of the form $m(t)= j_L(n(t))$ for a curve $n(t) \in N$, because the relation $(q,p)= \mathbb{F} L_S (q,v)$ is included in the system \eqref{LD_P} (see the third equation in \eqref{implicit_thermo_simple}). In addition, by inspection of the explicit forms \eqref{implicit_thermo_simple_4} and \eqref{Hamiltonian_side}, the curve $n(t)$ is solution of the Hamilton-Dirac system \eqref{HD_M}.

\begin{figure}[p]
\begin{center}
\includegraphics[scale=0.8]{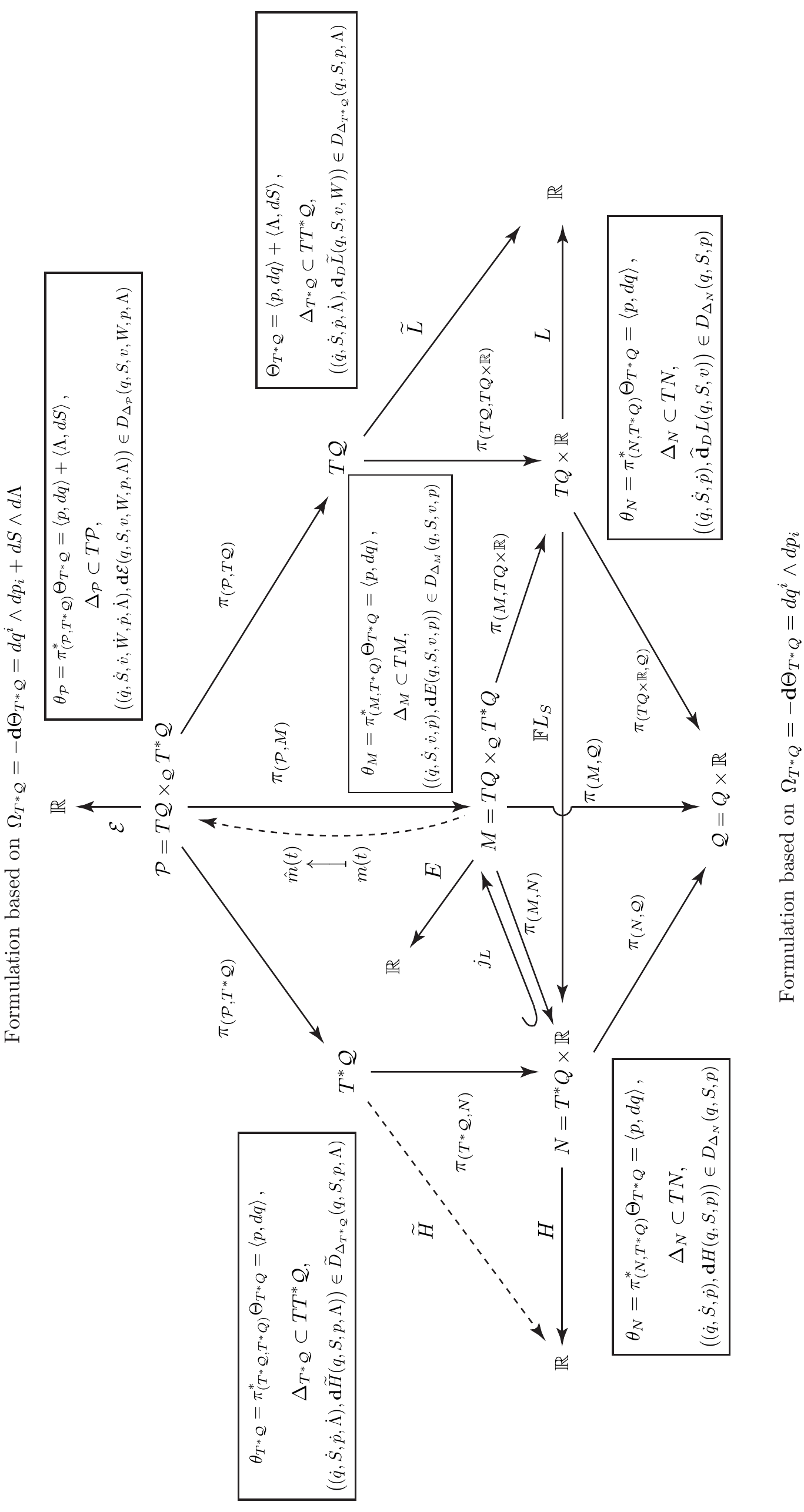}
\caption{Illustration of the relation between the Dirac formulations for the thermodynamics of simple systems developed in this paper. From the top to the bottom: the Dirac dynamical system formulation based on the generalized energy on $ \mathcal{P} $ developed in \S\ref{LD_Pontryagin_Therm};
the Lagrange-Dirac formulation on $T^* \mathcal{Q} $ developed in \S\ref{LD_Cotang_Therm};  the Dirac dynamical system formulation on $M$ developed in \S\ref{DF_4}; the Lagrange-Dirac and Hamilton-Dirac formulations developed in \S\ref{LD_4} and \S\ref{HD_4}.}
\label{big_diagram}
\end{center}
\end{figure}

\paragraph{Relation between the Dirac formulations on $M$ and $ \mathcal{P} $.} We note that if the curve $m (t) =(q(t),S(t),v(t),p(t)) \in M=TQ \oplus _ \mathcal{Q} T^*Q$ is a solution of the Dirac dynamical system \eqref{LD_system_4} on $M$, then the lifted curve $ \hat{m} (t) \in \mathcal{P} $ defined by
\[
\hat{m}(t):= \big( q(t),S(t),v(t),W(t)=\dot S(t), p(t), \Lambda (t)=0\big)
\]
is a solution of the Dirac dynamical system \eqref{LD_system} on $ \mathcal{P}$. See Fig. \ref{big_diagram} for an illustration of this relation.

\begin{remark}[On Hamilton-Dirac formulations]{\rm Note that we have two Lagrange-Dirac formulations: one on $N= T^*Q \times \mathbb{R}  $ which uses the Lagrangian $L:TQ \times\mathbb{R}  \rightarrow \mathbb{R}  $, see \S\ref{LD_4}, and one on $T^* \mathcal{Q} $ which uses the lifted Lagrangian $\widetilde{L}:T\mathcal{Q}\rightarrow  \mathbb{R}  $, see \S\ref{LD_Cotang_Therm}. On the Hamilton-Dirac side, however, we only have shown one formulation, on $N=T^*Q \times \mathbb{R}  $, described in the present section. There is no natural corresponding Hamilton-Dirac formulation on $T^* \mathcal{Q} $, because, in thermodynamics, the lifted Lagrangian $\widetilde{L}:T \mathcal{Q} \rightarrow \mathbb{R}  $ is always degenerate, since it does not depend on $\dot S$. 

Despite this degeneracy, one can define a Hamiltonian on $T^* \mathcal{Q} $ by lifting the Hamiltonian $H:T^*Q \times\mathbb{R}  \rightarrow \mathbb{R}  $ as follows
\[
\widetilde{H}:=H \circ \pi_{(T^{\ast}\mathcal{Q},N)}:T^* \mathcal{Q} \rightarrow \mathbb{R}.
\]
In this case, one recovers the equation of evolution for the thermodynamic of simple systems as a Hamilton-Dirac system for $\widetilde{H}$ on $T^* \mathcal{Q} $ by using a Dirac structure induced from the presymplectic form $ \omega _{T^* \mathcal{Q} }:= \pi _{( T^* \mathcal{Q} , T^*Q)} ^\ast \Omega _{T^*Q}$ (not the canonical symplectic form $ \Omega _{T^* \mathcal{Q} }$). Namely, we define
\begin{equation}\label{D_reduced}
\begin{aligned}
\widetilde{D}_{\Delta _{T^* \mathcal{Q} }}(z) :
&=\big\{ (v_{z}, \alpha_{z}) \in T_{z}T^*\mathcal{Q} \times T^{\ast}_{z}T^*\mathcal{Q} \; \mid \;  v_{z} \in \Delta _{T^* \mathcal{Q} }(z) \; \text{and} \\ 
& \qquad \qquad \qquad \left\langle \alpha_{z},w_{z} \right\rangle =\omega _{T^{\ast} \mathcal{Q} }(z)(v_{z},w_{z}) \; \;
\mbox{for all} \; \; w_{z} \in \Delta _{T^* \mathcal{Q} }(z) \big\},
\end{aligned}
\end{equation}
where $ \Delta _{T^* \mathcal{Q} }$ is the distribution defined in \eqref{loc_CVcot}. 
Note that the Dirac structure \eqref{D_reduced} is distinct from the Dirac structure defined in \eqref{D_thermo_qSpLambda}. The condition $\big((z, \dot z ),(z, \zeta  )\big) \in \widetilde{D}_{\Delta _{T^* \mathcal{Q} }}(z)$ is equivalent to
\[
\left\{
\begin{array}{l} 
\displaystyle\vspace{0.2cm}(\dot p+\alpha ) T(q,p,S)=    \mathcal{T}  \mathcal{F} ^{\rm fr}(q,p,S),\\
\displaystyle\vspace{0.2cm}T(q,p,S) \dot S= - \left\langle \mathcal{F} ^{\rm fr}(q,p,S), \dot q \right\rangle,\\
\displaystyle  u=\dot q, \quad \Psi  = 0.
\end{array}\right.
\]
Thus, the Hamilton-Dirac system
\[
\big((q,S,p, \Lambda ,\dot q, \dot S, \dot p, \dot \Lambda ),\mathbf{d} \widetilde{H}(q,S,p,\Lambda ) \big) \in \widetilde{D}_{T^*\mathcal{Q} }(q,S,p, \Lambda )
\]
does yield the system \eqref{local_thermo_HD4}. 
}
\end{remark} 


\begin{remark}[Inclusion of the external force]\label{Rmk_Fext}{\rm An external force $F^{\rm ext}: TQ \times \mathbb{R} \rightarrow T^* Q$ can be easily included in all the Dirac formulations. To do this, we consider the lift of the external force onto the manifold on which the Dirac structure is defined. For example, for the Dirac formulation on $ \mathcal{P} =T \mathcal{Q} \oplus T^* \mathcal{Q} $ developed in \S\ref{LD_Pontryagin_Therm}, we consider the projections $ \pi _{ (\mathcal{P} , TQ \times \mathbb{R}  )}:\mathcal{P} \rightarrow TQ\times \mathbb{R}  $ and $ \pi _ { (\mathcal{P} , Q)}: \mathcal{P} \rightarrow Q$, locally given by $  \pi _{ (\mathcal{P} , TQ \times \mathbb{R} ) }(q,S,v,W,p,\Lambda )= (q,v,S)$, $\pi _ {( \mathcal{P} , Q)}(q,S,v,W,p,\Lambda )=q$ and define the lifted external force as the horizontal one-form $F^{\rm ext}_ \mathcal{P} \in \Omega  ^1 ( \mathcal{P} )$ given by
\[
\left\langle F^{\rm ext}_ \mathcal{P} (x), v_x \right\rangle :=  \left\langle F^{\rm ext} \left(   \pi _{ (\mathcal{P} , TQ \times \mathbb{R} ) }(x) \right) , T  \pi _{ (\mathcal{P} , Q ) }( v_x) \right\rangle ,
\]
with $x= (q,S,v,W,p,\Lambda )$. This external force is included in the Dirac dynamical system \eqref{LD_system} as follows
\[
\big( (x, \dot x), \mathbf{d} \mathcal{E} (x)- F^{\rm ext}_ \mathcal{P} (x)  \big) \in D_{ \Delta  _ \mathcal{P} }(x).
\]
One easily checks that it yields the evolution equations \eqref{simple_systems}. The case of the other Dirac formulations is treated similarly.
}
\end{remark}

\section{Examples}\label{Examples} 

We illustrate the Dirac formulations with three examples of simple systems, involving three different irreversible processes: friction, matter transfer, and chemical reactions.

\paragraph{The one-cylinder problem.} This example has been already described in \S\ref{Sec_1}, see Fig. \ref{one_cylinder}. The thermodynamic configuration space is $ \mathcal{Q} = \mathbb{R}  \times\mathbb{R}  \ni (x,S)$. We shall illustrate the approach of \S\ref{DF_4}.  The bundle $M$ reads $ M= T \mathbb{R} \oplus  _{ \mathbb{R}  \times \mathbb{R}  } T^* \mathbb{R} \ni (x,S,v,p)$. The Dirac structure is given as follows. For each $(x,S,v,p ) \in M$, we have
\[
\big(( \dot x, \dot S, \dot v, \dot p ), (\alpha ,  \mathcal{T}  ,  \beta , u )\big) \in D_{ \Delta  _M} \;\;\Leftrightarrow\;\;
\left\{ 
\begin{array}{l}
\vspace{0.2cm}( \dot p+\alpha ) T(x,S)= -  \mathcal{T}  \lambda (x,S) v,\\
\vspace{0.2cm}T(x,S) \dot S= \lambda(x,S) v \dot x,\\
 \beta =0, \quad u=\dot x.
\end{array}
\right.
\] 
The generalized energy is
\[
E(x,S,v,p) = pv-\frac{1}{2} m v ^2+ U(x,S).
\]
and one directly checks that the Dirac dynamical system \eqref{LD_system_4} yields the evolution equations \eqref{equation_piston} for the piston. The other formulations can be carried out similarly, by using the expressions
\begin{align*} 
\mathcal{E} (x,S,v,W,p, \Lambda ) &= pv+ \Lambda W -\frac{1}{2} m v ^2+ U(x,S) \quad \text{on}\quad   \mathcal{P} = T (\mathbb{R}  \times\mathbb{R}  ) \oplus T^*( \mathbb{R}  \times \mathbb{R}  ),\\
H(x,S,p )&= \frac{1}{2m}p ^2 +U(x,S)  \quad \text{on} \quad N= T ^\ast \mathbb{R} \times \mathbb{R},  \\
\widetilde{H}(x,S,p, \Lambda )&= \frac{1}{2m} p ^2 +U(x,S) \quad \text{on}\quad  T^* \mathcal{Q} = T^*(\mathbb{R}  \times \mathbb{R}  ).
\end{align*}

\begin{figure}[h]
\begin{center}
\includegraphics[scale=0.55]{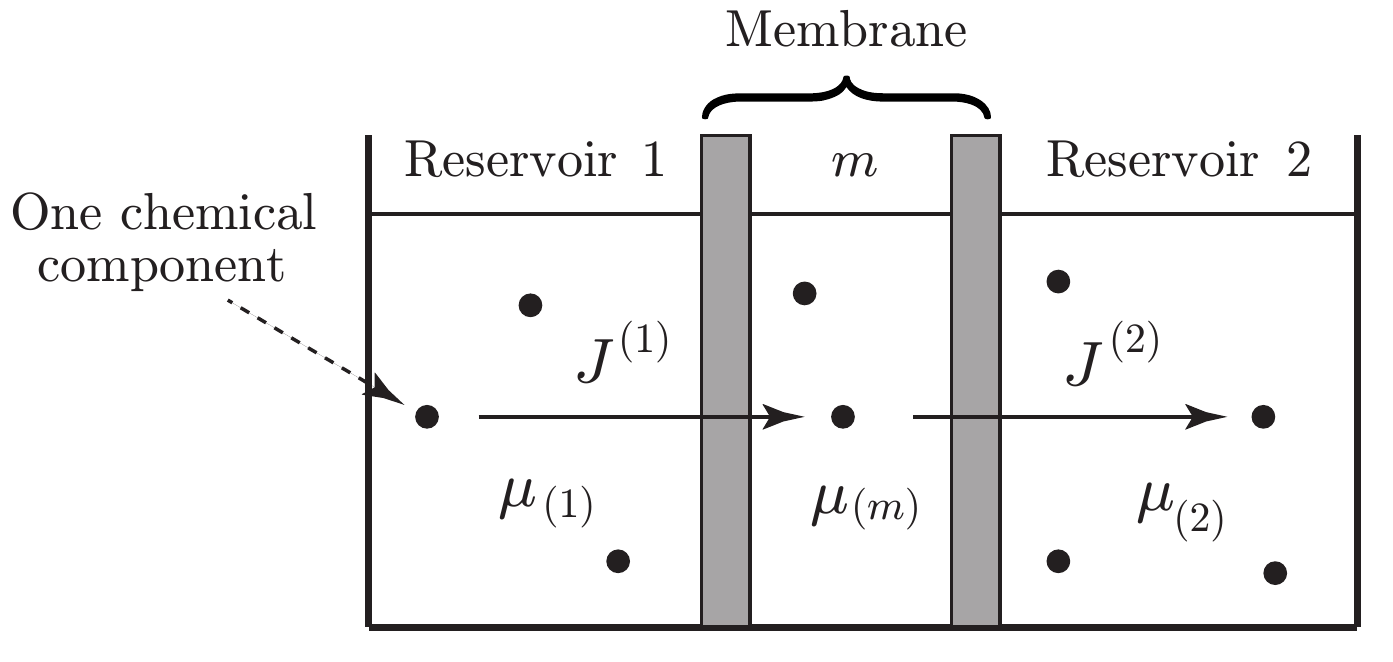}
\caption{Nonelectrolyte diffusion through a homogeneous membrane}
\label{MatterTransport}
\end{center}
\end{figure}

\paragraph{Diffusion through a homogeneous membrane.} We consider a {\it system with diffusion due to (internal) matter transfer} through a homogeneous membrane separating two reservoirs, as considered in \cite[\S2.2]{OsPeKa1973}. The equations of evolution where derived in \cite[\S3.4]{GBYo2016a} from the variational formalism, to which we refer for more details on this example. We suppose that the system is simple (so it is described by a single entropy variable) and involves a single chemical component. We denote by $N^{(m)}$ the number of mole of this chemical component in the membrane and also by $N^{(1)}$ and $N^{(2)}$ the numbers of mole in the reservoirs $1$ and $2$, as shown in Fig. \ref{MatterTransport}. 
\medskip

\noindent (1) \textit{Variational formulation based on the chemical potential.} The approach developed in \cite{GBYo2016a} is based on the internal energy $U(S, N^{(1)}, N^{(2)}, N^{(m)})$, expressed in terms of the entropy and number of moles. We shall here describe the system by using the thermodynamic potential expressed in terms of the entropy and the chemical potentials, namely, we define the thermodynamic potential
\[
\Phi (S,\mu _{(1)}, \mu _{(2)}, \mu _{(m)}):= U( S, N^{(1)}, N^{(2)}, N^{(m)})- \sum_{k=1,2,m} \mu _{(k)} N^{(k)},
\]
were $N^{(k)}$ on the right hand side are defined from the conditions $ \frac{\partial U}{\partial N^{(k)}}= \mu ^{(k)}$.

The Lagrangian in this representation is defined as
\begin{equation}\label{LPhi} 
L(w, \dot w, S):= - \Phi (S,\dot w_{(1)}, \dot w _{(2)}, \dot w _{(m)}),
\end{equation} 
where $w:=( w_{(1)}, w_{(2)}, w_{(m)})$.
Let us denote by $J^{(1)}$ the flux from the reservoir $1$ into the membrane and $J^{(2)}$ the flux from the membrane into the reservoir $2$. The variational formulation can be expressed as
\[
\delta \int_{t _1 }^{ t _2}L(w, \dot w , S)dt =0, \quad\;\;\; \textsc{Variational Condition}
\]
where the curves $w(t)$ and $S(t)$ satisfy the 
nonlinear nonholonomic constraint 
\[
\frac{\partial L}{\partial S}(q, \dot q, S)\dot S  =J^{(1)}( \dot  w_{(1)} - \dot  w_{(m)})  +J^{(2)}(\dot  w_{(m)} - \dot  w_{(2)}) \quad \textsc{Phenomenological Constraint}
\]
and with respect to variations $ \delta w(t) $ and $\delta S(t)$ subject to the constraint
\[
\frac{\partial L}{\partial S}(w, \dot w, S)\delta S= J^{(1)}( \delta  w_{(1)} - \delta  w_{(m)})  +J^{(2)}(\delta  w_{(m)} - \delta  w_{(2)}), \quad \textsc{Variational Constraint}
\]
with $ \delta w(t_1)=\delta w(t_2)=0$. This variational principle, together with the observation $ \frac{\partial L}{\partial \dot w^{(k)}}= - \frac{\partial \Phi  }{\partial \mu ^{(k)}} = N^{(k)}$ and the phenomenological constraint yield the evolution equations
\begin{equation}\label{membrane_equation} 
\left\{ 
\begin{array}{l}
\vspace{0.2cm}\dot N^{(1)} = J^{(1)},\quad \dot N ^{(m)} = -J^{(1)}+J^{(2)}, \quad \dot N ^{(2)} =-J^{(2)}\\
-T\dot S= J^{(1)}(\mu _{(1)}- \mu _{(m)})+ J^{(2)}( \mu _{(m)}- \mu _{(2)}).
\end{array}
\right.
\end{equation} 
\medskip

\noindent (2) \textit{Dirac formulation.} The thermodynamic configuration space is $ \mathcal{Q} = \mathbb{R}^3  \times\mathbb{R}  \ni (w,S)$ with $w=( w_{(1)}, w_{(2)}, w_{(m)})$. The bundle $M$ reads $M = T\mathbb{R} ^3  \oplus _{ \mathbb{R}  ^3 \times\mathbb{R}  } T^* \mathbb{R} ^3\ni (w,S,v, N )$, with $v=( w_{(1)}, w_{(2)}, w_{(m)})$, $N=(N^{(1)}, N^{(2)}, N^{(m)})$. The Dirac structure is given as follows. For each $(w,S,v,N ) \in M $, we have
\begin{align*} 
&\big(( \dot w, \dot S, \dot v,  \dot N ), (\alpha ,  \mathcal{T}  ,  \beta    , u )\big) \in D_{ \Delta _M}(w,S,v,N)\\
& \qquad \;\;\Leftrightarrow\;\;
\left\{ 
\begin{array}{l}
\vspace{0.2cm}( \dot N + \alpha ) T(w,S)=  \mathcal{T} \left( J^{(1)}( v_{(1)} - v_{(m)})  +J^{(2)}(v_{(m)} - v_{(2)}) \right) \\
\vspace{0.2cm}T(w,S) \dot S= J^{(1)}( \dot w_{(1)} - \dot w_{(m)})  +J^{(2)}(\dot w_{(m)} - \dot w_{(2)})\\
 \beta =0, \quad u=\dot w.
\end{array}
\right.
\end{align*} 
The generalized energy is
\[
E(w,S,v,N)= \sum_{k=1,2,m} N^{(k)} v_{(k)}+ \Phi (S, v_{(1)}, v _{(2)}, v _{(m)})
\]
and one directly checks that the Dirac dynamical system \eqref{LD_system_4} yields the evolution equation \eqref{membrane_equation} for membrane transport.

The generalized energy on $ \mathcal{P} =T( \mathbb{R}  ^3 \times\mathbb{R}  ) \oplus T^* (\mathbb{R} ^3 \times \mathbb{R}  )$ to be used in the Dirac formulation \eqref{LD_system} is
\[
\mathcal{E} (w,S,v,W,N, \Lambda ) =\sum_{k=1,2,m} N^{(k)} v_{(k)}+ \Lambda W+\Phi (S, v_{(1)}, v _{(2)}, v _{(m)}).
\]
\medskip

\noindent \textit{(3) The Hamiltonian.}  We note that the Lagrangian defined in \eqref{LPhi} is nondegenerate in the sense of \eqref{PLT}. The associate Hamiltonian defined in \eqref{def_H} recovers the internal energy, namely,
\[
H(w,N, S)= U( S, N^{(1)}, N^{(2)}, N^{(m)}).
\]
The Hamilton-Dirac system \eqref{HD_system_HD} yields the evolution equations \eqref{membrane_equation}.

\paragraph{Chemical reactions dynamics.} Consider a system of several chemical components undergoing chemical reactions. Let $I=1,...,R$ be the {\it chemical components}  and $a=1,...,r$ the chemical reactions. We denote by $N_I$ the {\it number of moles} of the component $I$. Chemical reactions may be represented by
\[
\sum_I {\nu '}_{I}^a\,I\; \stackrel[a _{(2)} ]{ a _{(1)} }{\rightleftarrows} \; \sum_I{\nu ''}^a_{I}\, I, \quad a=1,...,r,
\]
where $a_{(1)}$ and $a_{(2)}$ are {\it forward and backward reactions} associated to the reaction $a$, and ${\nu''}^a _{I}$, ${\nu '}^a_I$ are {\it forward and backward stoichiometric coefficients} for the component $I$ in the reaction $a$. From this relation, the number of moles $N_I$ has to satisfy
\begin{equation}\label{NH_constraint_reaction}  
\frac{d}{dt} N_I= \sum_{a=1}^{r} \nu^a _{I} \frac{d}{dt} \psi _a, \quad I=1,...,R,
\end{equation} 
where $\nu^a _{I}:= {\nu ''}^a_{I}- {\nu'}^a _{I}$, $ \psi _a $  is the {\it degree of advancement of reaction} $a$, and $\dot \psi _a $ is the {\it  rate of the chemical reaction} $a$. The mass conservation during each reaction is given by 
$$
\sum_I m_I\nu^a_{I}=0 \;\; \textrm{for $a=1,...,r$ (Lavoisier law)},
$$
where $m_I$ is the molecular weight of component $I$. We shall denote by $U=U(S, N_{1}, ..., N_{R},V)$, the internal energy of the system. We assume that the volume stay constant $V=V_0$.

The variational formulation for the nonequilibrium thermodynamics of this system was presented in \cite[\S3.3]{GBYo2016a} and is based on the Lagrangian defined by
\begin{equation}\label{L_CR} 
L( \psi, S):=-U(N_1,...,N_R, V_0, S),
\end{equation} 
where $ \psi =( \psi _1,..., \psi _r)$ and $N_I$ on the right hand side is expressed in terms of $ \psi $ by 
\begin{equation}\label{N_psi} 
N_I(t)=N_I(t_1)+ \sum_{a=1}^r\nu^a _{I} \psi  _a (t), \quad \psi_a  (t _1 )=0.
\end{equation} 
which results from \eqref{NH_constraint_reaction}. The expression of the entropy production involves friction forces of the general form $\mathcal{F}^{\rm fr\, a} (\psi , \dot \psi, S)=- \lambda ^{ab}(\psi_{a}, S)\dot \psi _b$, where the symmetric part of the matrix $ \lambda^{ab}$ is positive.

The thermodynamic configuration space is $ \mathcal{Q} = \mathbb{R} ^r  \times\mathbb{R}  \ni ( \psi ,S)$.   The bundle $M$ reads $M=T \mathbb{R}  ^r\oplus _{\mathbb{R}  ^r \times \mathbb{R}  } T^* \mathbb{R}  ^r \ni(\psi ,S,v,p)$, with $ \psi =(\psi _1,..., \psi _r )$, $v=( v_1,...,v_r)$, $p=(p ^1 , ..., p ^r )$.
The Dirac structure is given as follows. For each $( \psi ,S,v,p ) \in M $, we have
\[
\big(( \dot \psi , \dot S, \dot v,   \dot p ), (\alpha ,  \mathcal{T}  ,  \beta   , u )\big) \in D_{ \Delta _N}( \psi , S, v,p) \;\;\Leftrightarrow\;\;
\left\{ 
\begin{array}{l}
\vspace{0.2cm}( \dot p^b + \alpha ^b) T( \psi ,S)= -  \mathcal{T} \lambda ^{ab} ( \psi ,S) v_a\\
\vspace{0.2cm}T( \psi ,S) \dot S= \lambda^{ab}(x,S) v_a \dot \psi _b\\
 \beta =0, \quad u=\dot \psi .
\end{array}
\right.
\] 
The generalized energy is
\[
E ( \psi ,S,v,p ) = \sum_{a=1}^rp^av_a- L( \psi ,S)
\]
and one obtains from the Dirac dynamical system \eqref{LD_system_4} the evolution equations  for the chemical reactions, which coincide with those derived in \cite[\S3.3]{GBYo2016a}.
The generalized energy on $ \mathcal{P} = T( \mathbb{R}^r  \times \mathbb{R}  )\oplus T^*(\mathbb{R}  ^r \times \mathbb{R}  )$ to be used in the Dirac formulation \eqref{LD_system} 
\[
\mathcal{E} ( \psi ,S,v,W,p, \Lambda ) = \sum_{a=1}^rp^av_a+ \Lambda W- L( \psi ,S).
\]

The Lagrangian \eqref{L_CR} is degenerate hence the Hamiltonian $H:T^* \mathbb{R} ^r  \times \mathbb{R}  \rightarrow \mathbb{R}  $ cannot be defined.

One can alternatively describe chemical reaction dynamics by using a thermodynamic potential $\Phi $ analogous to the one introduced for matter transfer above. This alternative formulation is based on the approach for chemical reactions described in \cite[Def. 3.7]{GBYo2016a}, in which case the Lagrangian is regular and a Hamiltonian can be defined.

\paragraph{Conclusion.} In this paper we have shown that the equations of motion for nonequilibrium thermodynamics can be formulated in the context of Dirac structures that are associated with the Lagrangian variational setting developed in \cite{GBYo2016a,GBYo2016b}. We have considered the thermodynamics of simple systems, i.e., systems for which one entropy variable is sufficient to describe the irreversibility. We have proved that the equations of motion can be naturally formulated as \textit{Dirac dynamical systems} based on either the generalized energy, the Lagrangian, or the Hamiltonian. These formulations are associated to either the canonical symplectic form on the \textit{thermodynamic} phase space $T^* \mathcal{Q} $ (\S\ref{LD_formulation}) or the canonical symplectic form on the \textit{mechanical} phase space $T^*Q$ (\S\ref{section_4}). These formulations are compatible with the geometric formulation of classical mechanics given in terms of the canonical symplectic form, which is recovered in absence of the entropy variable and friction forces. We have explained the link between the various Dirac formulations and summarised it in Fig. \ref{big_diagram}. Finally we have illustrated these formulations with examples involving the irreversible processes of friction, matter transfer, and chemical reactions.

\paragraph{Acknowledgement.} F.G.B. is partially supported by the ANR project GEOMFLUID, ANR-14-CE23-0002-01; H.Y. is partially supported by JSPS Grant-in-Aid for Scientific Research (26400408, 16KT0024, 24224004), the MEXT ``Top Global University Project'' and Waseda University (SR 2017K-167).

\end{document}